\documentclass[aps,prb,twocolumn]{revtex4-2}
\usepackage{graphicx}
\usepackage{amsmath}
\usepackage{amssymb}
\usepackage{braket}
\usepackage{hyperref}
\usepackage{subfigure}
\usepackage[usenames,dvipsnames]{xcolor}
\usepackage{graphicx}
\usepackage{subfigure}
\usepackage[mathscr]{euscript}

\DeclareMathOperator{\Tr}{Tr}
\DeclareMathOperator{\Li}{Li}

\begin{document}

\title{ Detecting topological phase transitions through entanglement between disconnected partitions in a Kitaev chain with long-range interactions }
\author{Saikat Mondal}
\email{msaikat@iitk.ac.in}
\author{Souvik Bandyopadhyay}
\author{Sourav Bhattacharjee}
\author{Amit Dutta}

\affiliation{Department of Physics, Indian Institute of Technology Kanpur, Kanpur 208016, India}

\begin{abstract}
We explore the behaviour of the disconnected entanglement entropy (DEE) across the topological phases of a long-range interacting Kitaev chain where the long range interactions decay as a power law with an exponent $\alpha$.
 We show that while the DEE may not remain invariant deep within the topologically non-trivial phase when $\alpha<1$, it nevertheless shows a quantized discontinuous jump at the quantum critical point and can act as a strong marker for the detection of topological phase transition. We also study the time evolution of the DEE  after a sudden quench of the chemical potential within the same phase. In the short range limit of a finite chain, the DEE is expected to remain constant upto a critical time  after the quench, which diverges in the thermodynamic limit. However, no such critical time is  found to exist when the  long-range interactions dominate (i.e., $\alpha<1$).	 
\end{abstract}
\maketitle

\section{Introduction}\label{sec_intro}
In the past few decades, there has been a great advancement in the fields of topological phase transitions
which fall beyond the regime of the conventional Landau-Ginzberg theory~\cite{lutchyn10,alicea10,fulga11,lutchyn11}. These transitions are relevant to quantum information
processing~\cite{chen10,fidkowski10,cho17}.
Topological phase transitions between different phases occur when the bulk gap of the system closes. Kitaev model~\cite{kitaev01,sau12,degottardi11,thakurathi13,degottardi131,degottardi132,rajak14,dutta15,souvik20,souvik21} in one dimension is one such paradigmatic example hosting topological phases.\\

A Kitaev chain consists of non-interacting spinless fermions on a one dimensional lattice. The Hamiltonian of a short range Kitaev chain is characterized by the presence of nearest neighbour hopping and superconducting pairing terms, in addition to an on-site chemical potential at each fermionic site. Furthermore, the Kitaev chain is invariant under time-reversal, particle-hole and sub-lattice symmetries~\cite{souvik21}. The existence of these symmetries endows the system with a rich yet simple topological structure.
 The topology of a short-range interacting  Kitaev model~\cite{kitaev01,sau12,degottardi11,thakurathi13,degottardi131,degottardi132,rajak14,dutta15,souvik20,souvik21} is characterised by a topological  order parameter manifested in the form of an integer quantised winding number in periodic boundary conditions. A nontrivial bulk topology is reflected as symmetry-protected edge states (massless Majorana modes) in the corresponding  open boundary situation. Importantly, these edge states are robust against local and sufficiently weak perturbations which do not affect the bulk topology. As one crosses into the non-topological phase, the winding number vanishes and the edge states disappear. It is also interesting to note that Kitaev chain consisting of spinless fermions can be mapped to a transverse field XY model of half-integer spins using Jordan-Wigner transformations~\cite{lieb61,dutta15}.\\

The long range interacting Kitaev chain with power-law decaying superconducting pairing term has a much richer phase diagram~\cite{vodola14,vodola15,vodola16,viyuela16,lepori17,lepori2017,dutta17,utso18,smaity20,utso19,utso21} when the long range interaction dominates (i.e., the exponent characterizing spatial decay of the interaction  $\alpha<1$); a phase transition can occur between the phases with half-integer winding numbers~\cite{utso18,utso19,smaity20}, although the change of winding number across the critical point is still an integer. Furthermore, in the long-range case, the topologically non-trivial phase consists of massive Dirac modes, instead of massless Majorana modes. These massive Dirac modes arise due to the hybridization of massless Majorana modes in the presence of long range interactions. These topologically protected massive Dirac modes are also robust against local perturbations which do not exceed the energy gap between the topological modes and the bulk bands.

In recent times, it has been realized that the topological features of a given system are closely related to the entanglement properties of the ground state of the system. In this regard, various measures such as the  entanglement spectrum \cite{pollman10, fidkowski10,turner11}, topological entanglement entropy \cite{hamma05, levin06, kitaev06} and the disconnected entanglement entropy (DEE) ~\cite{zeng19,micallo20,fromholz20} have been proposed which are able to characterize various topological phases that a system may host. The DEE in particular, provides an ingenious way to detect topological phases by extracting out the entanglement present between the edge-localized modes in the non-trivial topological phases. However, it might be noted that by construction, the DEE is not a bulk topological invariant, even in short range systems. Since, the DEE for topological systems can be calculated without resorting to the momentum space, it can therefore serve as a robust topological marker, detecting the contribution of the edge states directly, even for systems lacking translational symmetry. Thus, the DEE acts similarly to an order parameter, acquiring a finite non-zero value in the non-trivial phase and vanishing in the trivial phase where no edge modes exist. In addition, a discontinuous jump in its value across the phases also helps to efficiently pinpoint the critical points in the system. Further, it has already been shown that in an infinitely long, short-range chain, the DEE remains invariant under a unitary evolution~\cite{micallo20}. 
However, the behaviour of the DEE in long-range systems under a unitary evolution has not been explored in earlier studies.

In this work, we explore the viability of the DEE to detect topological phases and phase transitions in the presence of long range interactions in the one-dimensional Kitaev chain. As we shall show, although the presence of long-range interactions can alter the behaviour of the DEE deep into the topological and trivial phases under certain circumstances, the DEE is still able to distinguish the different topological phases of the system and importantly, serve as a strong marker for the critical point separating inequivalent topological phases. Specifically, we observe that even in the presence of long range interactions, the DEE exhibits a quantized jump at the critical point separating inequivalent topological phases. However, unlike that of the short range interacting Kitaev chain, the DEE does not remain invariant in a unitary evolution following a sudden quench of a parameter of the system, even when the initial and the quenched systems are topologically equivalent.\\

This paper is organized as follows. A brief recapitulation of the topological properties of the Kitaev chain, both in the limit of short range and long range interactions, is provided in Sec.~\ref{sec_SRK}. In Sec.~\ref{sec_disconnected entangle entropy}, we investigate the DEE in the context of the long range interacting Kitaev chain and compare the results with that observed in the short range limit. The dynamical properties of the DEE under a unitary evolution are analysed in Sec.~\ref{sec_time_evolution_SD}. Concluding remarks are presented in the sec.~\ref{sec_conclusion}. We have provided a detailed discussion of the methods used in this paper in Appendix~\ref{sec_method}. In Appendix~\ref{sec_sharp}, a brief discussion on the sharpness of the jumps of the DEE at the critical points for the exponent $\alpha>1$ is provided. The discussion on the variation of the bulk contribution to the DEE with the exponent $\alpha$ for the long range interacting Kitaev chain is presented in Appendix~\ref{sec_bulk_sd}. Variation of the DEE with the length of the disconnected partition is provided in Appendix~\ref{sec_sd_ld}. 
We provide some comments on the variation of the DEE with the strength of the superconducting pairing for the short range as well as long range interacting Kitaev chain in Appendix~\ref{delta_dee}. In Appendix~\ref{app_H_eff}, we provide an analytical calculation showing the increase in the effective range of interaction of the effective Hamiltonian with the time following a quench. The dynamical properties of the DEE following sudden quenches of the strength of the superconducting pairing and the exponent $\alpha$ are analysed in Appendix~\ref{dee_time_delta}.
\begin{figure*}
	\centering
	\subfigure[]{
		\includegraphics[width=0.35\textwidth]{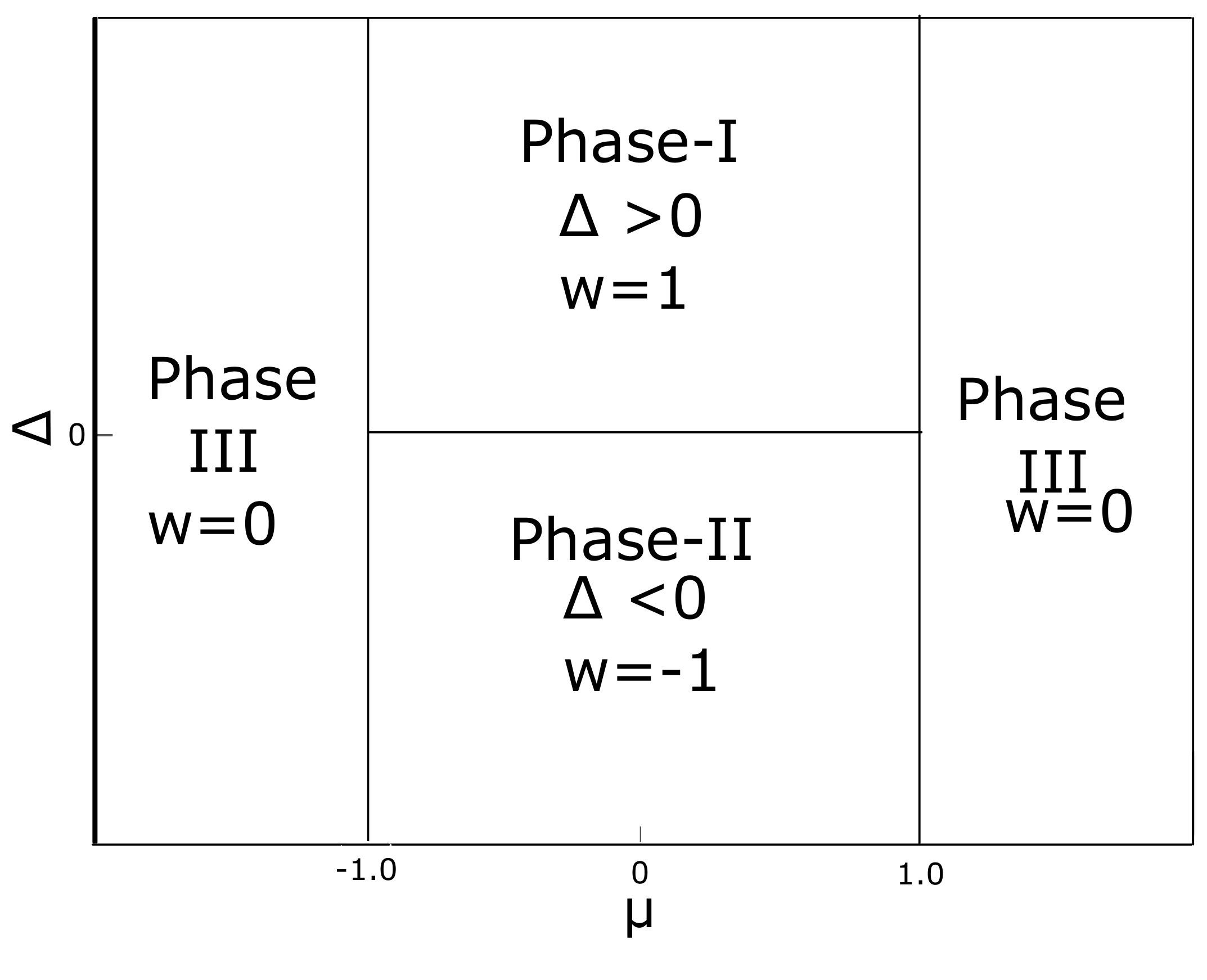}
		\label{fig_SRK_phase_diagram}}	
	\centering
	\hspace{15mm}
	\subfigure[]{
		\includegraphics[width=0.45\textwidth]{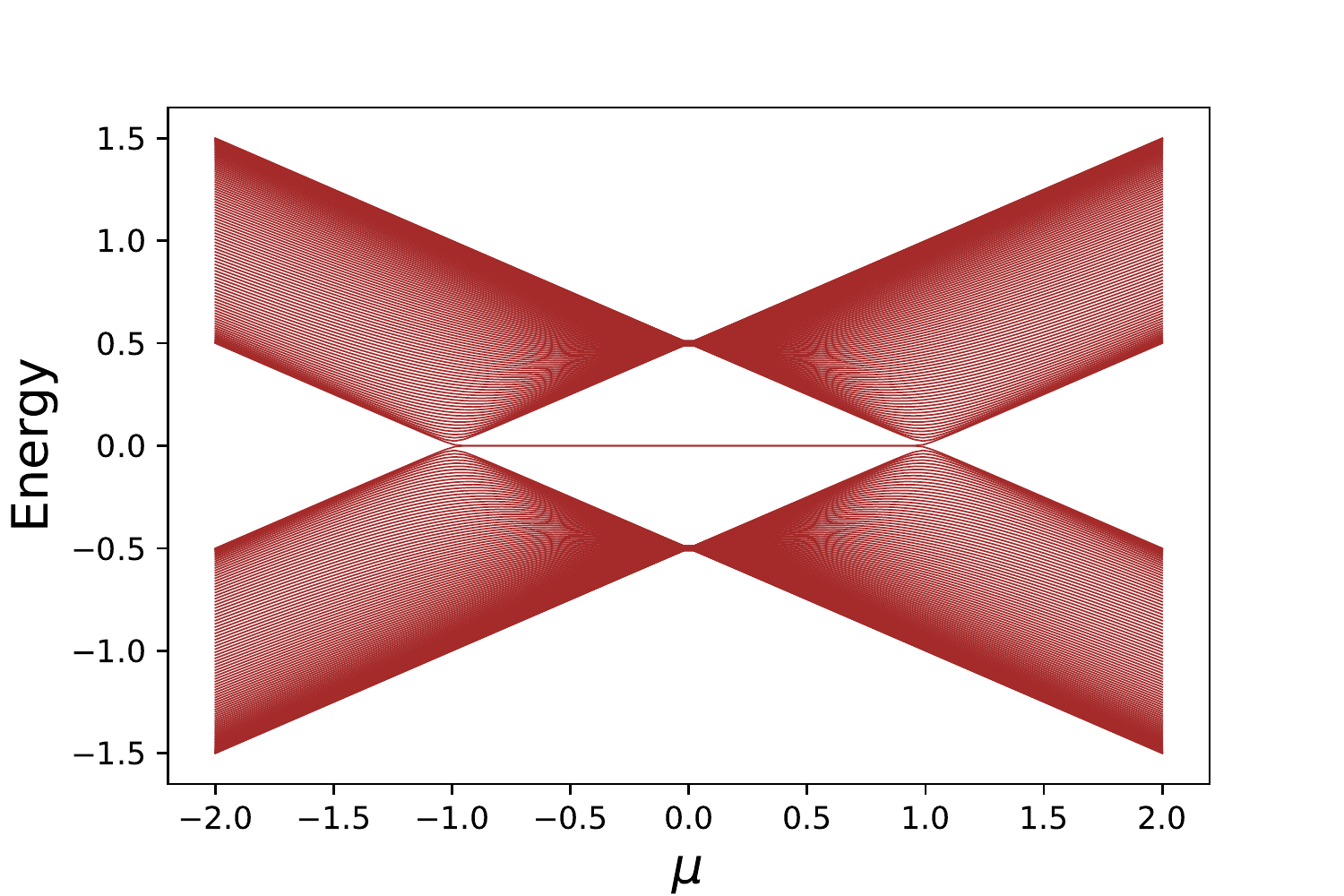}
		\label{fig_mu_E_short_range}}	
	\caption{(a) Phase diagram for short range interacting Kitaev chain ($\alpha\to\infty$). The phases with non-zero winding number $w$ are topologically non-trivial phases. (b) Single particle energy spectrum for the short range interacting Kitaev chain in open boundary condition with $\Delta=1$ and $L=100$. Zero energy Majorana modes exist in the topologically non-trivial phase (i.e., $-1<\mu<1$).}\label{fig_SRK}.
\end{figure*}

\section{Long range interacting Kitaev chain}\label{sec_SRK}
We consider the one-dimensional long range interacting Kitaev chain~\cite{vodola14,vodola15,vodola16,viyuela16,lepori17} with $L$ spinless fermions represented by the Hamiltonian,
\begin{multline}\label{eqn_LRK_H}
	H = \sum_{n=1}^{L-1} -\gamma( c_{n}^{\dagger} c_{n+1} + c_{n+1}^{\dagger} c_{n} ) - \mu \sum_{n=1}^{L} ( 2 c_{n}^{\dagger} c_{n} -1 ) \\ + \sum_{n=1}^{L-1} \sum_{l=1}^{L-n} \frac{\Delta}{l^{\alpha}} ( c_{n} c_{n+l} + c_{n+l}^{\dagger} c_{n}^{\dagger}),
\end{multline}
where $c_{n}$ ($c_{n}^{\dagger}$) is the annihilation (creation) operator of $n$-th fermionic site, $\Delta$ represents  the strength of the $p$-wave superconducting pairing interaction, $\mu$ is the on-site chemical potential. Henceforth,  we shall  set the nearest neighbour hopping parameter $\gamma=1$ everywhere.  Clearly, the exponent $\alpha(> 0)$ characterises the decay of  long-range interactions in the system.\\

Assuming periodic boundary conditions, the Hamiltonian in Eq.~\eqref{eqn_LRK_H} can be recast in Bogoliubov de-Gennes (BdG) basis as the following form :
\begin{equation}\label{eq_bdG}
H = \sum_{k \geq 0} \begin{pmatrix}
c_{k}^{\dagger} & c_{-k}
\end{pmatrix}  H_{k} \begin{pmatrix}
c_{k} \\
c_{-k}^{\dagger}
\end{pmatrix}.
\end{equation}
The Hamiltonian $H_{k}$ for $k\in[0,\pi]$ can be written as:
\begin{equation}\label{eq_H_22}
H_{k} = h_{y}(k) \sigma_{y} + h_{z}(k) \sigma_{z},
\end{equation}
where $\sigma_{y}$ and $\sigma_{z}$ are Pauli matrices. For long range interacting Kitaev chain, we have, $h_{y} (k) = \Delta f_{\alpha}(k)$ and $h_{z}(k) = \gamma \cos(k) + \mu$, where $f_{\alpha}(k) = \frac{i}{2} \left( \Li_{\alpha} (e^{-i k}) - \Li_{\alpha} (e^{i k}) \right)$. Here, the function $\Li_{\alpha}(x) = \sum_{l=1}^{\infty} \frac{x^{l}}{l^{\alpha}}$ is the poly-logarithmic function of $x$. Therefore, the dispersion relation with momentum $k\in[0,\pi]$ for long range interacting Kitaev chain takes the form,
\begin{equation}\label{eq_disp_lrk}
E_{k} = \sqrt{(\gamma \cos(k) + \mu)^2+ (\Delta f_{\alpha}(k))^2}.
\end{equation}
It is also important to note that in the short-range limit (i.e., $\alpha \to \infty$), we recover, $f_{\infty}(k)=\sin(k)$ being characteristic to the short range Kitaev Hamiltonian. \\

The various topological phases of the Kitaev chain are characterized by different values of the winding number ($w$)~\cite{utso18,utso19,smaity20}, which is defined as follows (see appendix of Ref.~\cite{smaity20} for a detailed calculation of winding number in a long range interacting Kitaev chain):
\begin{equation}\label{eq_winding}
w = \frac{1}{2 \pi} \int_{-\pi}^{\pi} dk \frac{d \phi_{k}}{dk},
\end{equation}
where $\phi_{k} = \tan^{-1} \left(\frac{h_{y}(k)}{h_{z}(k)}\right)$.
\\

Details of the methods used in this paper are provided in Appendix~\ref{sec_method}.

\subsection{Short-range limit ($\alpha>1$)}
In the limit $\alpha \to \infty $, the Hamiltonian in Eq.~\eqref{eqn_LRK_H} reduces to that of the short range Kitaev chain with only nearest-neighbour interactions~\cite{vodola14,lepori17,smaity20}. In this limit, the model hosts the following phases in the ground state:
\begin{enumerate}
\item \textit{Topologically non-trivial phase}: This phase lies within the parameter regimes $-1 < \mu < 1$ with $\Delta > 0 $ (phase-I) or $\Delta < 0$ (phase-II), as shown in Fig.~\ref{fig_SRK_phase_diagram}. The non-triviality of this phase is reflected in the non-zero integer quantized value of the winding number ($w=\pm 1$) \cite{thakurathi13,rajak14,souvik21}. The physically discernible feature of this phase is the manifestation of localized  Majorana zero modes (see Fig.~\ref{fig_mu_E_short_range}) at the edges of the open chain. Importantly, the Majorana zero modes at the opposite edges are long-range entangled.\\

\item {Topologically trivial phase}: This consists of phase-III ($ |\mu| > 1$). Majorana zero mode does not exist in the topologically trivial phase. The winding number in this phase is $w=0$ \cite{thakurathi13,rajak14,souvik21}.\\
\end{enumerate}

For  $\alpha > 1$, the phase diagram and the topological properties of the long-range Kitaev chain are identical to that of a short range Kitaev chain (i.e., $\alpha\to\infty$)~\cite{vodola14,utso19,smaity20}. However, as $\alpha$ approaches $1$, the bulk gradually starts becoming gapped near $\mu=-1$ and for $\alpha<1$, $\mu=-1$ no longer remains a critical point.

\begin{figure*}
	\centering
	\subfigure[]{
		\includegraphics[width=0.35\textwidth]{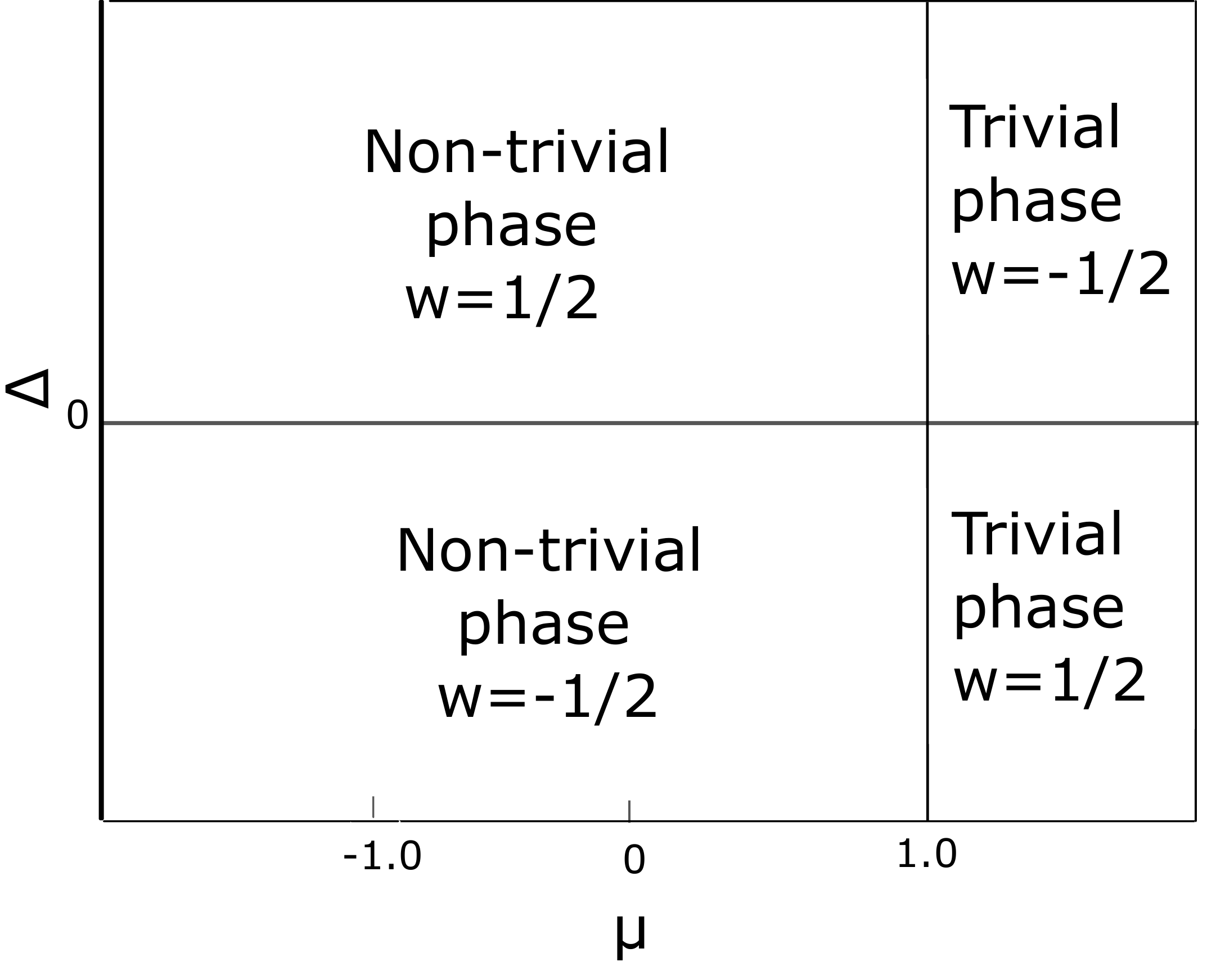}
		\label{fig_LRK_phase_diagram}}	
	\centering
	\hspace{15mm}
	\subfigure[]{
		\includegraphics[width=0.45\textwidth]{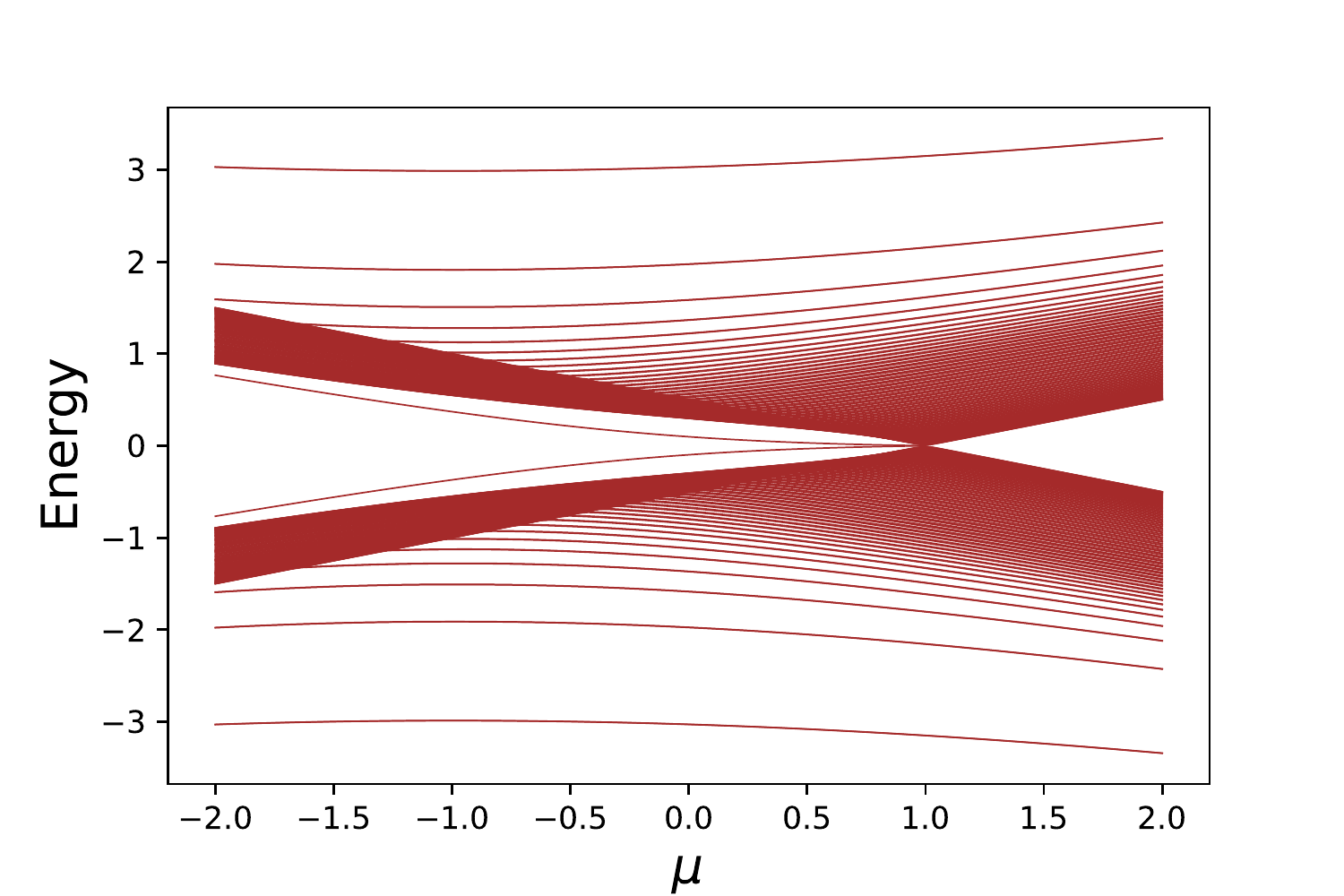}
		\label{fig_mu_E_alpha_0.5}}	
	\caption{(a) Phase diagram for long range interacting Kitaev chain with $\alpha<1$. Both the trivial and non-trivial phases have non-zero winding numbers ($w$). The signs of the winding numbers are flipped when the sign of $\Delta$ is flipped in both the phases. (b) Single-particle energy spectrum for long range interacting Kitaev chain in open boundary condition with $\alpha=0.5$, $\Delta=1$ and $L=100$. Massive Dirac modes exist in the non-trivial phase (i.e., $\mu < 1$).} 
\end{figure*}

\subsection{Long-range limit ($\alpha<1$)}\label{subsec_kitaev_long}
The topological properties of the Kitaev chain become markedly different when $\alpha < 1$ \cite{vodola14,utso19,smaity20}, as shown in Fig.~\ref{fig_LRK_phase_diagram}.  The system exists in a non-trivial phase for $\mu < 1$ and it is characterized by a half-integer quantized winding number $w=1/2$ ($w=-1/2$) when $\Delta>0$ ($\Delta<0$)~\cite{smaity20}. Due to the long range interaction, two Majorana zero modes at the two edges combine to form massive Dirac modes in this phase \cite{viyuela16,utso19,smaity20} as shown in Fig.~\ref{fig_mu_E_alpha_0.5}. On the contrary, the system remains topologically trivial for $\mu>1$. As there is no edge state in the trivial phase ($\mu>1$) for $\alpha<1$, the winding number ($w$) is expected to be zero in this phase. However, unlike the case of $\alpha>1$, the winding number is half-integer quantized in the topologically trivial phase, although the sign of the winding number flips across the critical point $\mu=1$. The half integer value of the winding number in the trivial phase indicates the weakening of the bulk-edge correspondence in the long-range interacting topological systems~\cite{lepori2017}. \\

\begin{figure}
	\centering
	\includegraphics[width=0.4\textwidth]{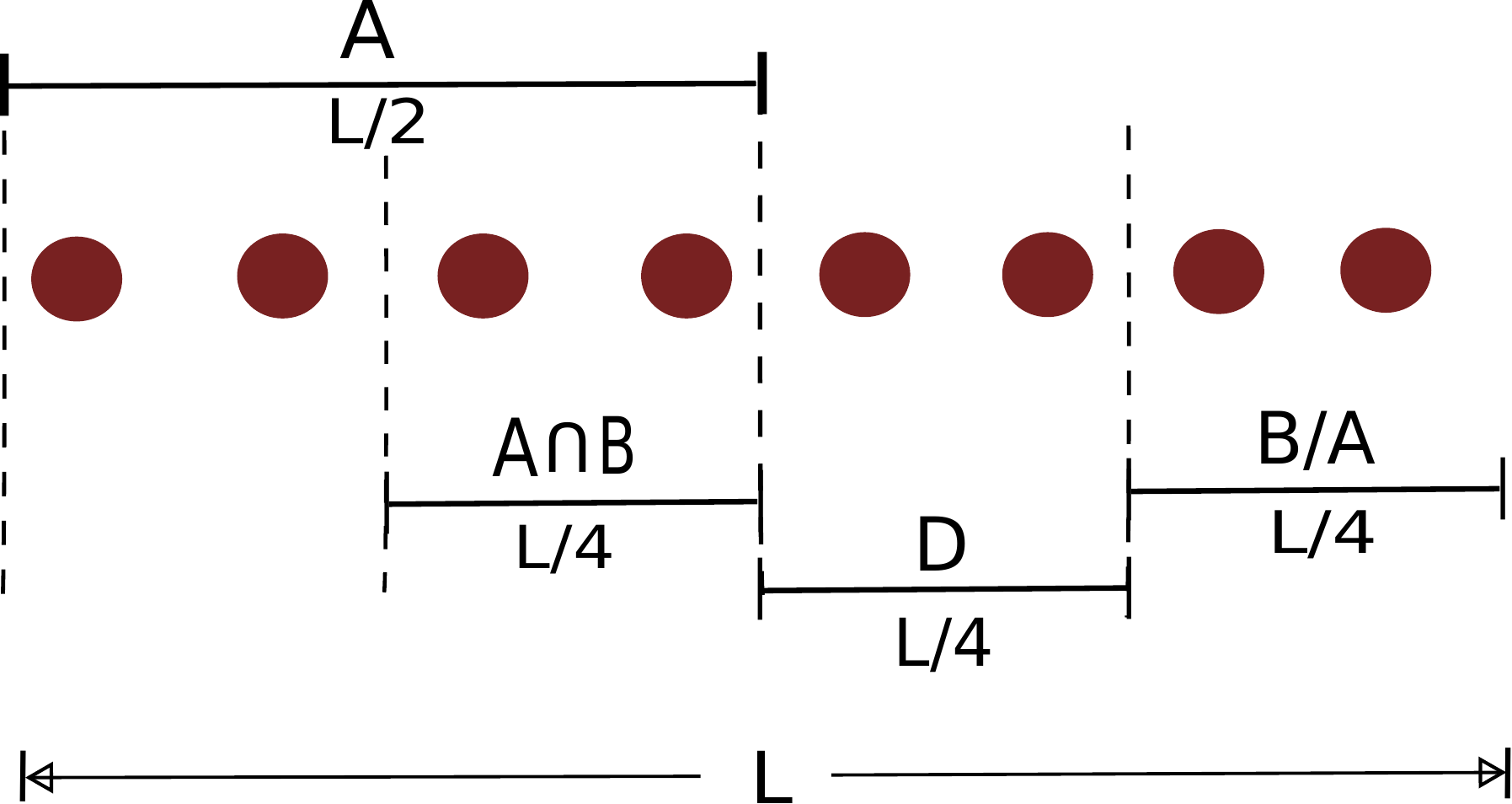}
	\centering
	\caption{Partitioning scheme chosen to calculate the disconnected entanglement entropy where $D = \overline{A \cup B}$ and $2 L_{A} = 2 L_{B} = 4 L_{D} = L$}\label{fig_partitions}
\end{figure}

\begin{figure*}
	\centering
	\subfigure[]{
		\includegraphics[width=0.45\textwidth]{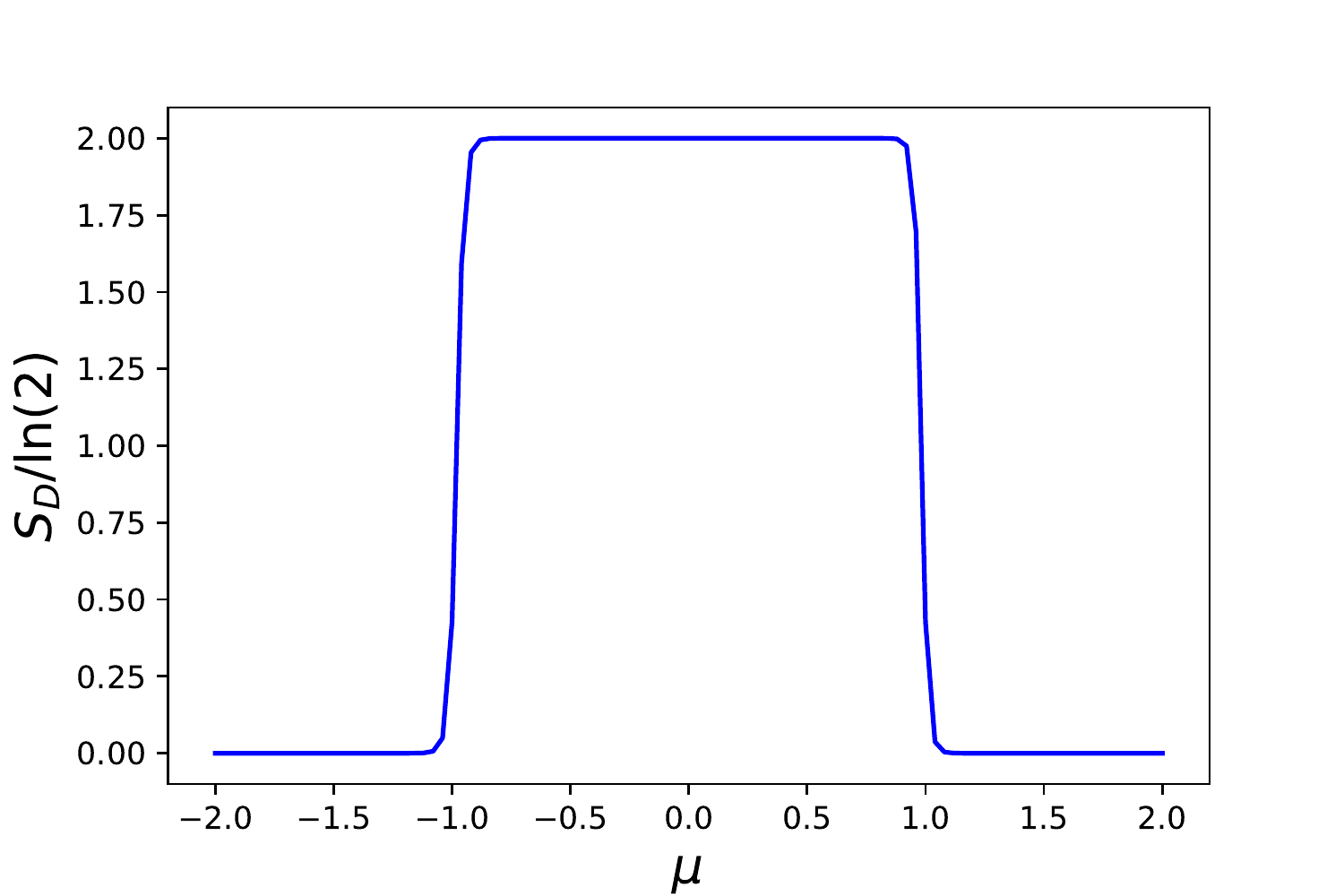}
		\label{fig_mu_SD_short_range}}	
	\centering
	\hspace{10mm}
	\subfigure[]{
		\includegraphics[width=0.45\textwidth]{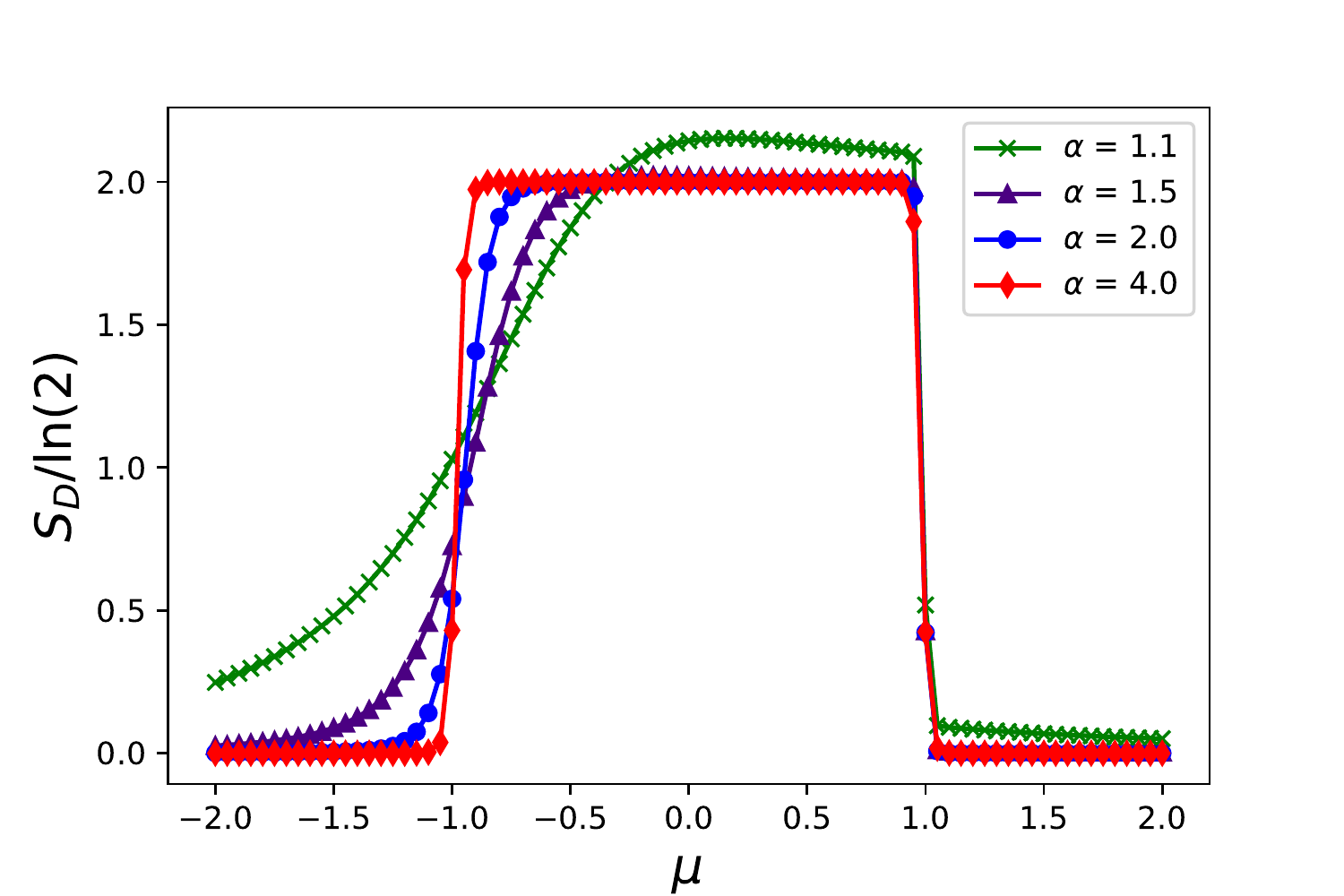}
		\label{fig_mu_SD_alpha_2.0}}
	\caption{The DEE as a function of the chemical potential in the (a) short range limit and (b) long range scenario with different values of $\alpha>1$. In both the cases, the DEE is able to separate the trivial phases from the non-trivial ones and shows jumps at the critical points. The other relevant parameters chosen are $\Delta=1$ and  $L = 100$. For $1 <\alpha \leq 2$, the jump of the DEE at $\mu=-1$ is less sharp compared to that at $\mu=1$.}
\end{figure*}


\section{Disconnected entanglement entropy}\label{sec_disconnected entangle entropy}
For a system existing in a pure state described by a density matrix $\rho$, the bipartite entanglement entropy \cite{latorre03,calabrese04,latorre09,calabrese09} of a part $X$ of the system is given by,
\begin{equation}\label{eqn_SA}
	S_{X} = - \Tr_{X} (\rho_{X} \ln (\rho_{X}) ), 
\end{equation}
where $\rho_X$ is the reduced density matrix of the subsystem $X$ obtained by partial trace over the degrees of freedom of the rest of the system, i.e., $\rho_X=\Tr_{\bar{X}}\rho$. Further, it is easy to show that $S_X=S_{\bar{X}}$.\\

Given an one-dimensional system of $L$ sites, let us now partition the system  as shown in Fig.~\ref{fig_partitions}. The two subsystems $A$ and $B$ overlap with each other over a finite region. Note that the partition $B$ is not continuous and spans on either side of a disconnected region $D = \overline{A \cup B}$. The disconnected entanglement entropy~\cite{micallo20,fromholz20} is then defined as, 
\begin{equation}\label{eqn_SD}
	S_{D} = S_{A} + S_{B} - S_{A \cup B} - S_{A \cap B} .
\end{equation}
As we shall elaborate below, the DEE, by construction, is able to extract the entanglement contribution arising from the presence of localised edge states which are long-range entangled in nature. Therefore, the DEE turns out to be an excellent probe to detect topological phases in in short-range interacting systems~\cite{micallo20} where the bulk states are short-range entangled. In Sec.~\ref{sec:lrk}, we shall investigate if the sensitivity of the DEE is hampered in presence of inherent long-range interactions in the system.\\  

Throughout this paper, the DEE has been calculated numerically using Eq.~\eqref{eqn_SD}, where the Von Neumann entropies $S_{X}$'s for $X=A,B,A \cap B, A \cup B$ have been evaluated from the eigenvalues of correlation matrices~\cite{latorre03,latorre09} having the elements $C_{ij}=\langle c_{i}^{\dagger} c_{j} \rangle$ and $F_{ij}=\langle c_{i} c_{j} \rangle$, in the ground state. (For more details, see Appendix~\ref{sec_method}.)

\subsection{DEE in the limit $\alpha\to\infty$}

Let us first consider the behaviour of the DEE in the case of the short-range interacting Kitaev chain with nearest-neighbour interactions (i.e., $\alpha\to\infty$). From Fig.~\ref{fig_mu_SD_short_range}, it is clear that the DEE assumes the values,
\begin{equation}\label{eqn_SD_SRK}
	S_{D} = \begin{cases}
		2 \ln (2) , & \text{for $-1 < \mu < 1$},  \\
		0 , & \text{for $|\mu| > 1$},
	\end{cases}
\end{equation}
thus acting as an order parameter distinguishing the topologically non-trivial phase from the trivial ones. To understand the above behaviour, we first note that no long range entanglement exists in the ground state of Kitaev chain in the trivial phase when $\alpha\to\infty$. For sufficiently large $L_D$, it therefore follows that,
\begin{equation}\label{eqn_SB_bulk}
 	S_{B} = S_{A \cap B}+ S_{B/A},
\end{equation}
as partitions $A\cap B$ and $B/A$ are separated by the disconnected partition $D$. Similarly, we also have,
 \begin{equation}\label{eqn_S_A_u_B}
 	S_{A \cup B} = S_{A} + S_{B/A}.
 \end{equation}
Substituting Eqs.~\eqref{eqn_SB_bulk} and~\eqref{eqn_S_A_u_B} in Eq.~\eqref{eqn_SD}, it is straightforward to see that $S_{D} = 0$ in the trivial phase, i.e., when $|\mu|>1$. \\

However, in the topologically non-trivial phase, the presence of long range entangled Majorana edge modes immediately implies that Eqs.~\eqref{eqn_SB_bulk} and~\eqref{eqn_S_A_u_B} are no longer satisfied. In fact, one can determine the finite value acquired by $S_D$ in this case by explicitly calculating the contribution to the DEE from the edge modes, i.e.,
\begin{equation}\label{eqn_SD_edge}
S_{D} = S^{lm} +S^{rm} ,
\end{equation}
where, $S^{lm}$ and $S^{rm}$ are the contributions of left and right localized edge modes to the DEE, respectively. 
As there are two possible microstates corresponding to each edge state, we therefore have $S^{lm} = S^{rm} = \ln (2) $ . Thus, in the topologically non-trivial phase, $S_{D} = 2 \ln (2)$.


\begin{figure*}
	\centering
	\subfigure[]{
		\includegraphics[width=0.45\textwidth]{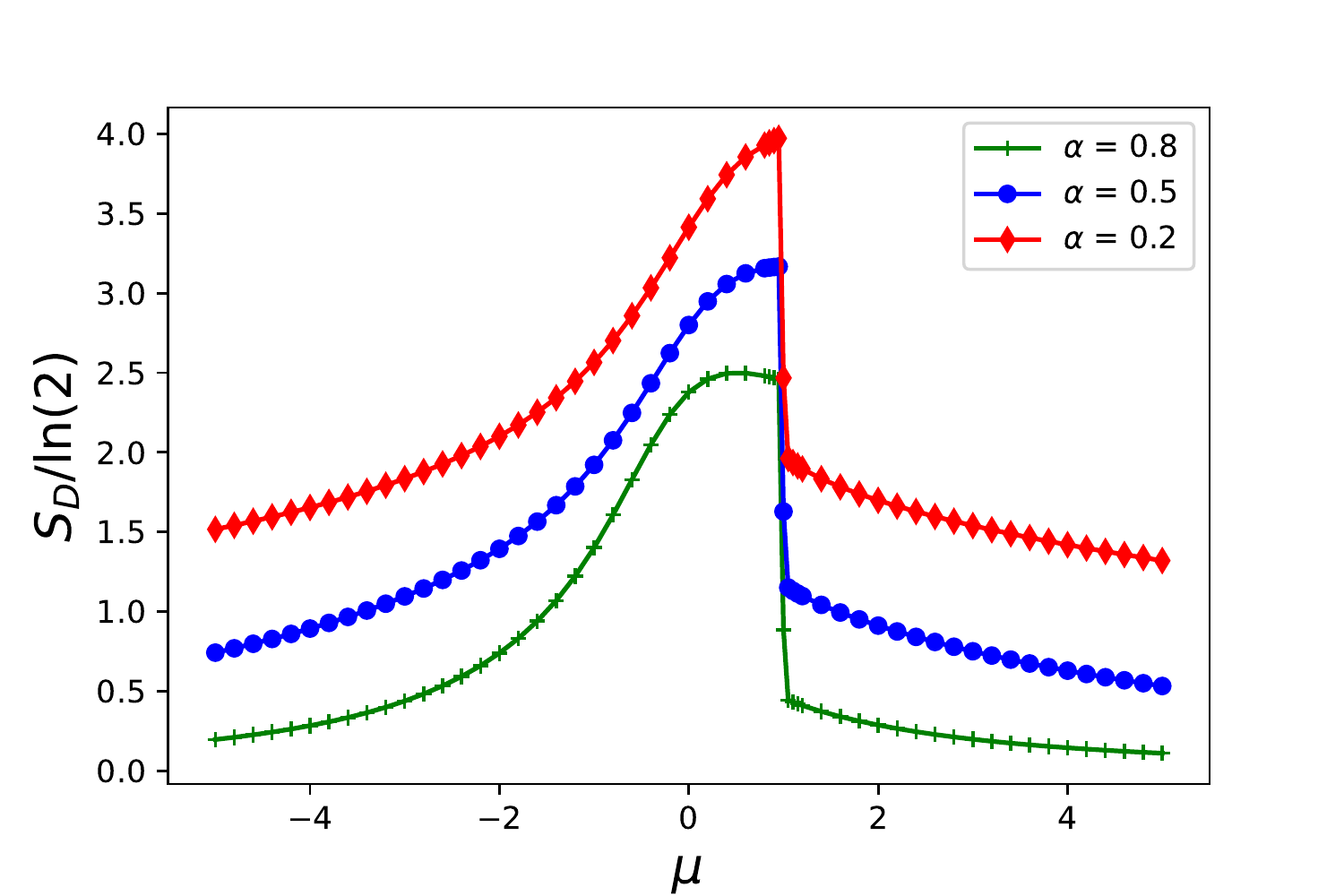}
		\label{fig_mu_SD_alpha_dependence}}	
	\centering
	\subfigure[]{
		\includegraphics[width=0.45\textwidth]{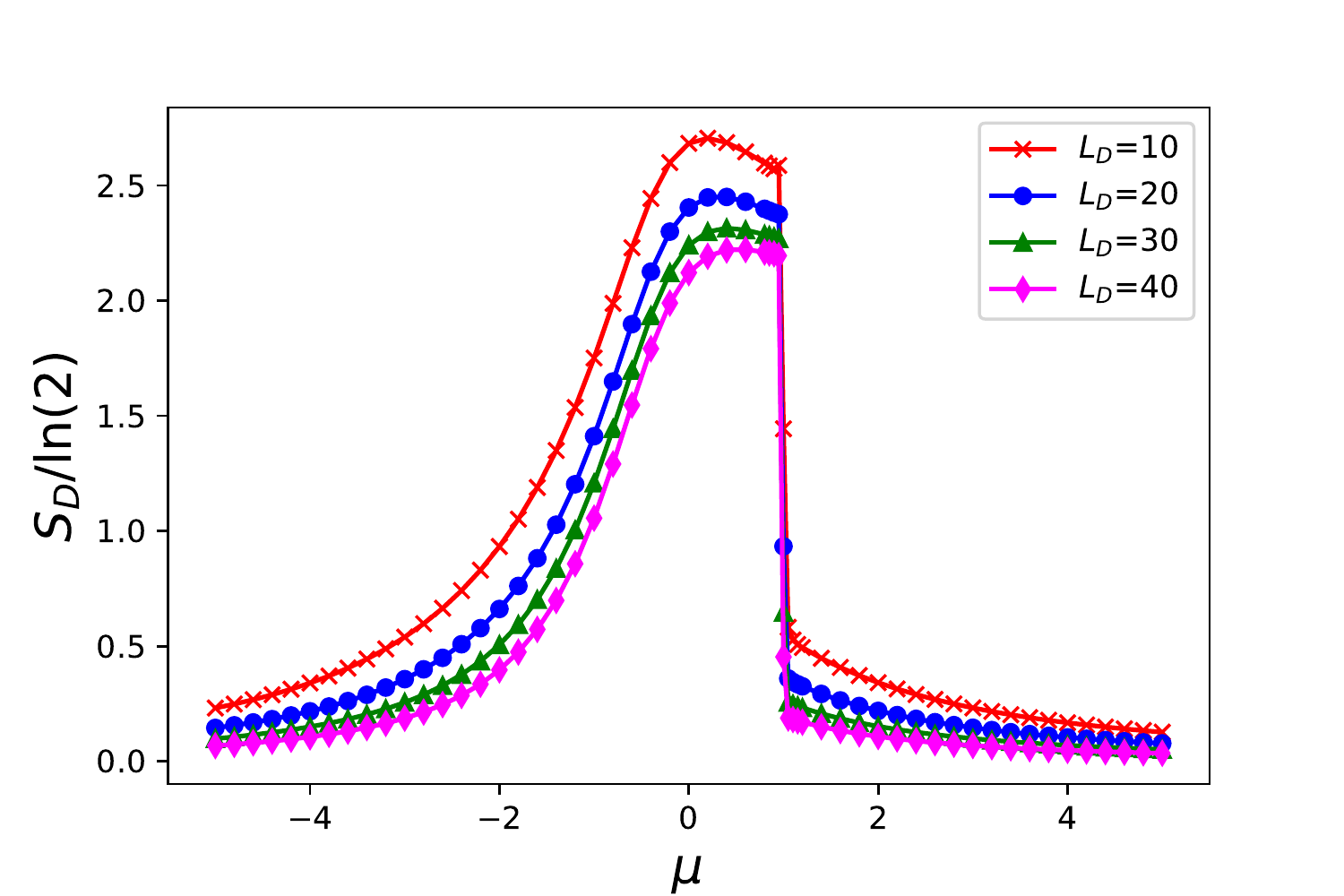}
		\label{fig_mu_SD_LD_dependence}}
	\caption{(a) The DEE as a function of the chemical potential in the long range interacting Kitaev chain with $L_D=25$, shown for different values of $\alpha$. (b) $S_D$ for different lengths ($L_{D}$) of the disconnected partition with fixed $\alpha=0.9$. The other relevant parameters chosen for the plots are $\Delta=1$ and $L = 100$. }\label{fig_mu_SD_alpha_LD}.
\end{figure*}


\subsection{DEE for a finite $\alpha$}
\label{sec:lrk}
\begin{enumerate}
\item \textbf{DEE for $ \boldsymbol{\alpha>1}$} : 
We have discussed in Sec.~\ref{subsec_kitaev_long} that the topological properties of the long-range Kitaev chain for any finite $\alpha>1$ are identical to that observed in the short-range limit, $\alpha\to\infty$. In Fig.~\ref{fig_mu_SD_alpha_2.0}, we show that this similarity is also reflected in the behaviour of the DEE. Despite the presence of long-range interactions, the DEE is clearly able to distinguish the non-trivial phase from the trivial phase. Further, it acquires the same finite value of $2\ln (2)$ in the non-trivial phase. \\

However, for $1<\alpha \leq 2$, the jump of the DEE at the critical point $\mu=-1$ does not show a sharp discontinuity as compared to that at $\mu=1$, in finite size systems. An explanation of this based on the scaling of the bulk gaps at the critical points with the system size ($L$) is provided in Appendix~\ref{sec_sharp}. 


\item \textbf{DEE for $\boldsymbol{\alpha<1}$} : 
The situation is trickier in the case of $\alpha < 1$, as shown in Fig.~\ref{fig_mu_SD_alpha_dependence}. In this case, the system has only one critical point at $\mu=1$. We find that the criticality is manifested as a discontinuity in the DEE; the latter however does not remain invariant deep in the non-trivial phase $\mu\ll 1$. This happens as the the topological edge states hybridize to form massive Dirac modes, thereby dispersing into the bulk deep into the non-trivial phase (see Fig.~\ref{fig_mu_E_alpha_0.5}).  Furthermore, the DEE does not immediately vanish in the trivial phase as $\mu$ is increased beyond the critical point $\mu=1$. This is because, even in the topologically trivial phase, the DEE contains non-zero contributions from long range entangled bulk states with no topological origin, a typical property of long range interacting systems. Therefore, the DEE can no longer be considered as a bulk topological invariant {\it per se}, as it is not quantized in long range interacting systems. We remark that, the failure of the bulk-boundary correspondence also manifests while characterizing topological phases in long range systems through conventional bulk topological invariants \cite{lepori2017}.\\

For $\alpha<1$, the dominant long-range interactions in the system leads to the generation of long-range entanglement in the bulk states themselves. In other words, the bulk entanglement generated by virtue of the long-range interactions implies that the additive decompositions in Eqs.~\eqref{eqn_SB_bulk} and~\eqref{eqn_S_A_u_B} are are not generically possible, unless when $L_D\to\infty$. When $\alpha$ is decreased, keeping $\mu$ fixed, the range of interaction increases, which leads to a greater value of the DEE in the non-trivial phase, since with long-range interactions the bulk contribution to the DEE increases. A brief discussion on the variation of the bulk contribution to the DEE with $\alpha$ is provided in Appendix~\ref{sec_bulk_sd}. The variation of the DEE with $L_{D}$ for long range interacting Kitaev chain is discussed in Appendix.~\ref{sec_sd_ld}. We observe that the DEE approaches $2\ln(2)$ and zero as the critical point $\mu=1$ is approached from the topologically non-trivial and trivial phase, respectively, as $L_{D} \to\infty$ for a thermodynamically large system.\\

Although in this section we have discussed the variation of the DEE with the chemical potential ($\mu$), it is also interesting to explore the variation of the DEE with the strength of the superconducting pairing ($\Delta$) for both short-range and long-range interacting Kitaev chain. A brief discussion of that is provided in Appendix~\ref{delta_dee}, where we have shown that the DEE exhibits a dip at $\Delta=0$ for the short-range as well as the long-range interacting systems.
\end{enumerate}

\section{Time evolution of disconnected entanglement entropy}\label{sec_time_evolution_SD}
Let us now consider a system with an arbitrary range of interaction given by the Hamiltonian $H_{i}$. At $t=0$,  the system is suddenly quenched, such that the Hamiltonian of the system becomes $H_{f}$, which also has an arbitrary range of interaction but is topologically equivalent to $H_i$. If the system is initially prepared in an eigenstate $\ket{\psi(0)}$ of the Hamiltonian $H_{i}$, then at any time $t>0$, the state of the system is determined by
\begin{equation}\label{eqn_psi_t}
\ket{\psi(t)}=U(t) \ket{\psi(0)} ,
\end{equation}
with $U(t)=\exp(-i H_{f} t)$. We define a time dependent, effective Hamiltonian~\cite{micallo20}
\begin{equation}\label{eqn_H_eff}
H_{\rm eff}(t)=U(t) H_{i} U^{\dagger}(t) ,
\end{equation}
such that $\ket{\psi(0)}$ and $\ket{\psi(t)}$ are the eigenstates of the Hamiltonians $H_{i}$ and $H_{\rm eff}(t)$ respectively, with the same energy eigenvalue. Using the Baker-Campbell-Hausdorff formula, the Eq.~\eqref{eqn_H_eff} can be recast as,
\begin{equation}\label{eqn_H_eff_series}
H_{\rm eff}(t)= H_{i} + \sum_{r=1}^{\infty} \frac{(-i t)^{r}}{r\,!} K_{r} (H_{f},H_{i}) ,
\end{equation}
where $K_{r} (H_{f},H_{i})=[H_{f},[H_{f},[....,[H_{f},H_{i}]....]]]$, with the Hamiltonian $H_{f}$ appearing $r$-times (see Ref.~\cite{micallo20}). {For the specific case in which both $H_i$ and $H_f$ are short-ranged, following straight forward algebraic simplification as demonstrated in {Appendix \ref{app_H_eff}}, we show that although $H_{\rm eff}$ is a finite ranged Hamiltonian, its range of interaction $r_{\rm max}(t)$ increases with time.
Now, if the maximum range of interaction of $H_{\rm eff}$ spans the complete disconnected partition $D$ or greater (see Fig.~\ref{fig_partitions}), bulk states start becoming entangled over regions exceeding the disconnected partition size. This starts introducing bulk corrections to the DEE as seen for long ranged interacting Hamiltonians, which in turn breaks the temporal invariance of the DEE. Hence, a finite disconnected partition size gives rise to a critical time scale in short-range Hamiltonians until which the DEE remains temporally invariant under all unitary driving generated by a short range Hamiltonian topologically equivalent to $H_i$. We therefore  define a critical time $t_{c}$, such that $r_{\rm max}(t_{c}) \sim L_{D}$. However if we one sets $L_D\propto L$, for example $L_D\sim fL$ such that $f<1$, both $L_D$ and the critical maximum interaction range ($r_{\rm max}(t_c)$) of $H_{\rm eff}$ become infinitely large in the thermodynamic limit. Assuming a finite velocity of the increasing range of $H_{\rm eff}$, one might then infer that the critical time $t_c$ also grows linearly in system size $L$ (as can be concluded from Fig.~\ref{fig_t_SD_LD}). It can therefore be expected that the disconnected entanglement entropy ($S_D$) for short range interacting Kitaev chain, at all times $t < t_{c}$, should have a constant value of either $2 \ln(2)$ or $0$, if $H_{i}$ and $H_{f}$ are topologically equivalent.\\

\begin{figure*}
	\centering
	\subfigure[]{
		\includegraphics[width=0.45\textwidth]{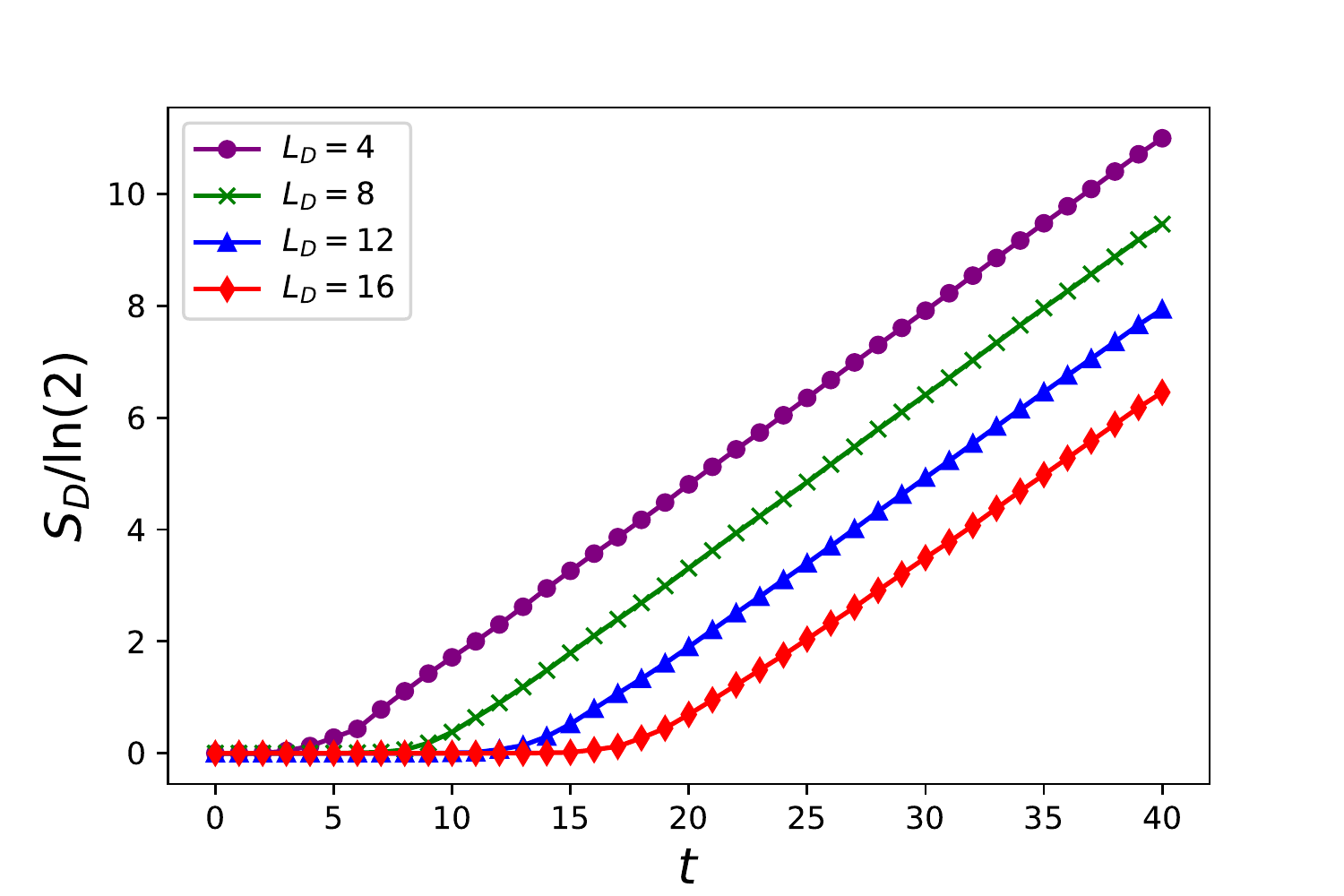}
		\label{fig_t_SD_LD}}	
	\centering
	\subfigure[]{
		\includegraphics[width=0.45\textwidth]{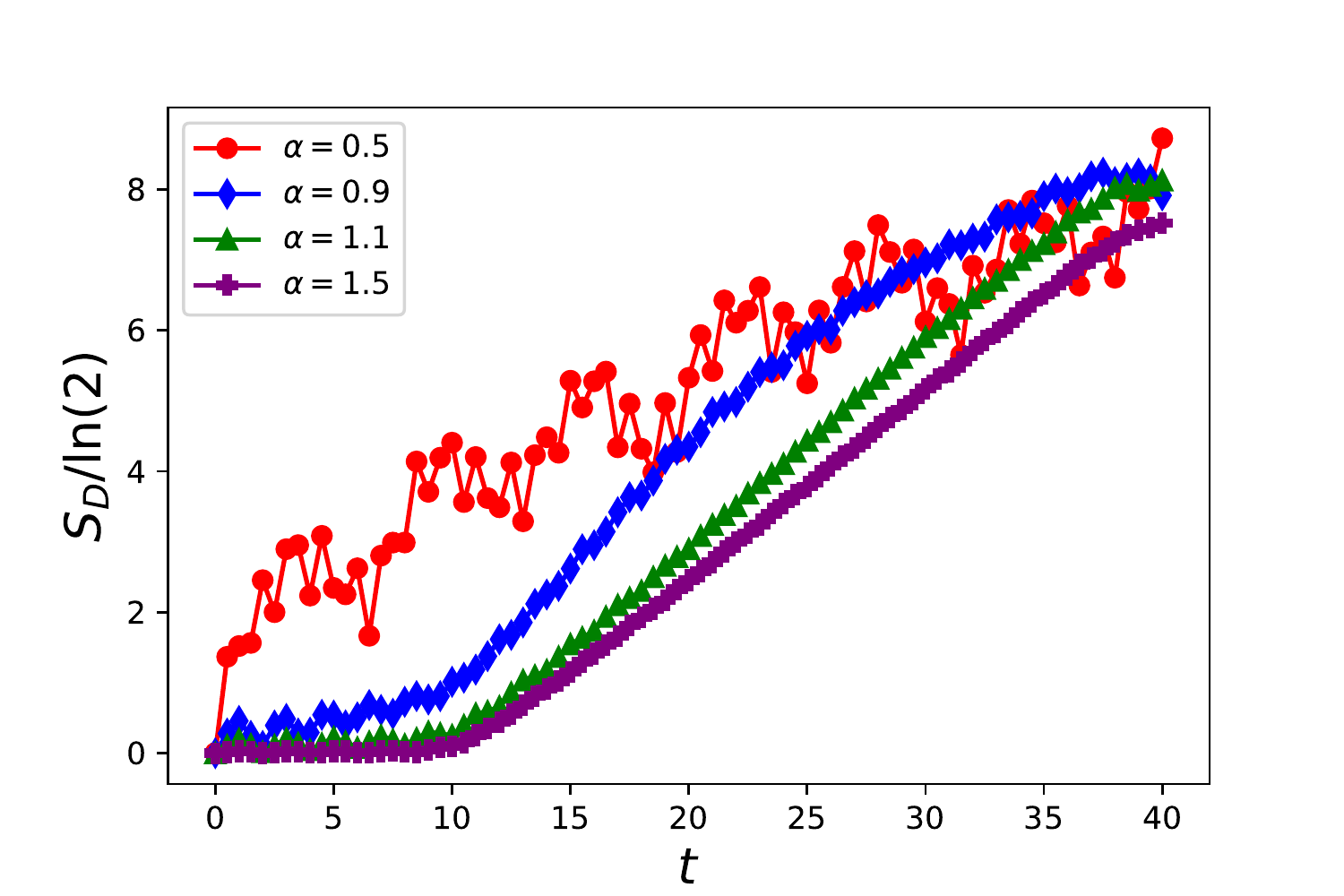}
		\label{fig_t_SD_alpha}}
	\caption{(a) $S_{D}$ as a function of time ($t$) for different values of $L_{D}$ for short range interacting Kitaev chain with $L=40$, $\mu_{i}=100$ and $\mu_{f} = 2$, $\Delta = 1$, (b) $S_{D}$ as a function of time for different values of $\alpha$ for long range interacting Kitaev chain with $L=40$, $L_{D}=10$, $\mu_{i}=100$ and $\mu_{f}=2$, $\Delta = 1$. Here, $\mu_{i}$ and $\mu_{f}$ denote the chemical potentials before and after the quench, respectively.} \label{fig_t_SD}.
\end{figure*}

For the long range interacting Kitaev chain, as the interactions appearing in the Hamiltonian $H_{i}$ are already long ranged and $H_{\rm eff}(t=0) = H_{i}$, we have, $r_{\rm max}(t=0)\sim L$. Thus, it is expected that the critical time for long range interacting Kitaev chain is $t_{c} \sim 0$ and $S_{D}$ starts to change with time from $t \sim 0$ itself. This suggests that in contrast to short ranged systems, the topological information captured by $S_D$ does not remain stable under unitary dynamics. As illustrated in Fig.~\ref{fig_t_SD_LD} and Fig.~\ref{fig_t_SD_alpha}, $t_{c}$ is indeed zero for long range interacting ($\alpha < 1$) Kitaev chain and is proportional to $L_{D}$ (for constant $L$) for short range interacting Kitaev chain, respectively. We also observe from the Fig.~\ref{fig_t_SD_LD} that $S_{D}=0$, $\forall t < t_{c}$ for short range interacting Kitaev chain as both $H_i$ and $H_f$ are chosen to be topologically trivial.\\

Although the time evolution of the DEE following a sudden quench of the parameter $\mu$ has been discussed in this section, it is also possible to observe the same behaviour of the DEE following a sudden quench of the strength of superconducting pairing ($\Delta$). In Appendix~\ref{dee_time_delta}, we have provided the behaviour of the DEE in a Kitaev chain of finite size following sudden quenches of the strength of superconducting pairing ($\Delta$) and the exponent $\alpha$ of the power law interaction.

\section{Conclusion}\label{sec_conclusion}
 For a short range interacting Kitaev chain, in a topologically non-trivial phase, only the edge states contribute to the disconnected entanglement entropy ($S_{D}$). In general, if there are $p$ edge states in the topologically non-trivial phase of a short range interacting system, then the number of possible microstates in the space of the topological edge states (with respect to their occupations) is $2^{p}$ and the disconnected entanglement entropy, $S_{D}= \ln(2^{p})= p \ln{2}$ \cite{micallo20}. As there is no edge localized state in topologically trivial phase, disconnected entanglement entropy in this phase is zero. As $S_{D}$ remains invariant under the continuous change of the parameter $\mu$ in the same topological phase in the short range interacting systems, the disconnected entanglement entropy ($S_{D}$) can be thought of a topological invariant for short range interacting systems. As $S_{D}$ exhibits discontinuous jumps at the quantum critical points (i.e., $\mu=1$ and $\mu=-1$), it can be used as a marker for the detection of topological phase transitions.\\

On the contrary, in the long range interacting Kitaev chain, both the bulk and edge states contribute to $S_{D}$ due to the presence of long-range correlated bulk states in addition to the massive edge states. Therefore, in the topologically non-trivial phase of long range interacting system, $S_{D}$ turns out to be greater than that for short range interacting system due to the contribution of the bulk states to $S_{D}$. Also, $S_{D}$ is non zero in the topologically trivial phase (i.e., $\mu >1$) of long range interacting Kitaev chain, due to non zero contribution of bulk states. As $S_{D}$ does not remain constant with the change of $\mu$ in the same topological phase in the long range interacting Kitaev chain, it can not be considered as strong topological invariant for long range interacting Kitaev chain in the conventional sense. However, we show that $S_{D}$ exhibits a discontinuous jump at the critical point (i.e., $\mu=1$) of long range interacting Kitaev chain which is also quantized in $\ln2$. Thus, it can still be used to sharply detect a topological phase transition even in the long range interacting systems.\\

Similar to Ref.~\cite{micallo20}, we  also observe that under a unitary evolution generated by a quenched Hamiltonian which is topologically equivalent to the initial system, the time evolution of $S_{D}$ shows the existence of a critical time scale $t_{c} \neq 0$ in the short range interacting Kitaev chain, upto which $S_D$ remains invariant. This in turn suggests that for time $t < t_{c}$, $S_{D}$ remains invariant in time and starts to change only after $t=t_{c}$. This is explained by the fact, that the effective interactions of the system remain short ranged with respect to the length of the disconnected partition and $S_{D}$ remains a good topological invariant for $t<t_{c}$. At $t \geq t_{c}$, the effective interactions of the system span the complete disconnected partition length and therefore, $S_{D}$ no longer remains a good topological invariant due to the inclusion of bulk contributions. Thus, $S_{D}$ starts to change with time for $t\geq t_{c}$ for short range interacting systems. However, for long range interacting system, as the effective interactions are long ranged (spanning the complete system size and hence the disconnected partition size) starting from $t=0$ itself. We conclude that, in contrast to short range interacting systems, the disconnected entanglement entropy does not remain stable under unitary evolution for long range interacting systems. That is, $S_{D}$ starts varying with time immediately after the quench at $t=0$ for long range interacting Kitaev chain.

Given the efficacy of the the entanglement between disconnected partitions in detecting topological phase transitions in both short range and long range interacting systems, it will be interesting to probe the thermalization of topological properties in such systems in the presence of integrability perturbations. For example, the quantity $S_D$ has already been seen to effectively capture topological phase transitions in the presence of local disorder in short range interacting systems (see Ref.\cite{micallo20}). Furthermore, in short range interacting one-dimensional topological systems, it has been seen that the dynamical or explicit breaking of protective symmetries lead to a time varying topological invariant which in turn connects to the flow of a electric polarisation in the out of equilibrium system \cite{cooper18,duttasouvik19,souvik19}. In this regard, it might also be important to study the behaviour of the DEE under dynamical and explicit symmetry breaking and connect its dynamics to a macroscopic polarisation current in the system. Another direction of research might be to generalise the disconnected entanglement entropy in more than one dimensional systems to characterize symmetry protected and Chern topological phases in higher dimensions.

\begin{acknowledgements}
	S.M. and S. Bandyopadhyay acknowledge support from PMRF fellowship, MHRD, India. S. Bhattacharjee acknowledges CSIR, India for financial support. 
	A.D. acknowledges financial support from SPARC program, MHRD, India and
	SERB, DST, New Delhi, India. We acknowledge Somnath Maity for comments.
\end{acknowledgements}

\appendix

\section{Details of the methods used}\label{sec_method}
We define Majorana operators ($a$'s) in the following way:
\begin{subequations}\label{eq_majorana}
\begin{equation}\label{eq_majorana1}
a_{2n-1}=c_{n}^{\dagger}+c_{n} ,
\end{equation}
\begin{equation}\label{eq_majorana2}
a_{2n}=\frac{1}{i} (c_{n}^{\dagger}-c_{n}) ,
\end{equation}
\end{subequations}
for $n=1,2,...,L$. The Majorana operators satisfy the anti-commutation relation: $\{a_{m},a_{n}\}=2 \delta_{mn}$. The Hamiltonian in Eq.~\eqref{eqn_LRK_H} of the main text can be recast as the following equation in terms of Majorana operators:
\begin{align}\label{eq_m_H}
H= - \frac{i \gamma}{2} \sum_{n=1}^{L-1} (a_{2n}a_{2n+1}- a_{2n-1}a_{2n+2}) + i\mu \sum_{n=1}^{L} a_{2n-1}a_{2n} \nonumber \\
- \sum_{n=1}^{L-1} \sum_{l=1}^{L-n} \frac{i \Delta}{2 l^{\alpha}} (a_{2n-1}a_{2n+2l}+ a_{2n}a_{2n+2l-1}).
\end{align}
Using the anti-commutation relations of Majorana operators, Eq.~\eqref{eq_m_H} can be further written in the Majorana basis as:
\begin{equation}\label{eqn_M}
H=\sum_{m=1}^{2L} \sum_{n=1}^{2L} a_{m} M_{mn} a_{n} ,
\end{equation}
with $M$ being a $2L \times 2L$ antisymmetric matrix. From the eigenvalues of the matrix $M$, single particle energy spectrum is obtained.\\

Now, Von Neumann entropies $S_{X}$ for the partitions $X=A,B,A\cap B$ and $A\cup B$ are evaluated from the correlation matrix. For any partition $X$ with $l$ fermionic sites, we define the $2l \times 2l$ correlation matrix as,
\begin{equation}\label{eq_corr_m}
\mathcal{C} = \begin{pmatrix}
I - C & F \\
F^{\dagger} & C
\end{pmatrix} ,
\end{equation}
where $I$ is the $l \times l$ identity matrix, both $C$ and $F$ are $l \times l$ matrices with the elements $C_{ij}=\langle c_{i}^{\dagger} c_{j} \rangle$ and $F_{ij}=\langle c_{i} c_{j} \rangle$ respectively, with $i,j=1,2,...,l$, calculated in the ground state of the chain. If $\lambda_{i}$'s are the eigenvalues of the correlation matrix $\mathcal{C}$, then the Von Neumann entropy $S_{X}$ can be written as,
\begin{equation}\label{eq_entropy}
S_{X}= -\sum_{i=1}^{2l} \lambda_{i} \ln(\lambda_{i}) .
\end{equation}

\section{Sharpness of the jumps of the DEE at the critical points for $\alpha>1$}\label{sec_sharp}
For $1<\alpha \leq 2$, the jump of the DEE at the critical point $\mu=-1$ does not show a sharp discontinuity as compared to that at $\mu=1$, in systems with finite size. This can be explained from the dispersion relation in Eq.~\eqref{eq_disp_lrk} of the main text. The critical values of the momentum $k$, for which the bulk energy gaps at the critical points $\mu=-1$ and $\mu=1$ vanish, are $k=0$ and $k=\pi$, respectively. Expanding $E_{k}$ about $k=0$ at the critical point $\mu=-1$ ~\cite{vodola15}, it can be shown that for any $\alpha \neq 2$,
\begin{equation}\label{eq_disp_mu_n}
E_{k \to 0} = \sum\limits_{n=1}^{\infty}B_{n}(\alpha) k^{n} + A(\alpha) k^{\alpha-1},
\end{equation}
where $A$ and $B_n$ are functions of $\alpha$. From Eq.~\eqref{eq_disp_mu_n}, it can be observed that the leading order terms contributing to $E_{k \to 0}$ at $\mu=-1$ are different for $\alpha>2$ and $\alpha<2$, which are given by,
\begin{equation}\label{eq_disp_zero_case}
	E_{k \to 0} \sim  \begin{cases}
		k  ,& \text{for $\alpha>2$},  \\
		k^{\alpha-1}  ,& \text{for $\alpha<2$}.
	\end{cases}
\end{equation}
For the marginal case $\alpha=2$, the leading order dependence of $E_{k \to 0}$ on the momentum $k$ turns out to be,
\begin{equation}\label{eq_disp_alpha2}
	E_{k \to 0} \sim C_{1} k + C_{2} k \ln(k),
\end{equation}
where $C_{1}$ and $C_{2}$ are constants. In finite size systems, the Brillouin zone is discrete and the lower momentum cutoff near the critical point $\mu=-1$ scales as $L^{-1}$. Therefore, from Eq.~\eqref{eq_disp_zero_case} it can be observed that the corresponding bulk gap at $\mu=-1$ vanishes as $1/L$ for $\alpha>2$ and as $1/L^{\alpha -1}$ for $\alpha<2$.\\

On the other hand, at the critical point $\mu=1$ the leading order term of $E_{k}$ for $k \to \pi$ is always linear,
\begin{equation}\label{eq_pi}
E_{k \to \pi} \sim \left| k- \pi \right|.
\end{equation} 
 Thus, the bulk gap at $\mu=1$, vanishes as $1/L$ for any finite value of $\alpha$. As the bulk gaps at both the critical points $\mu=1$ and $\mu=-1$ for $\alpha>2$ vanish as $L^{-1}$, the finiteness of the system takes it equally away from criticality in both the cases. This is reflected in the identical sharp jumps at $\mu=1$ and $\mu=-1$, for $\alpha>2$.\\

On the contrary, as evident from Eq.~\eqref{eq_disp_zero_case}, for $1<\alpha \leq 2$ the bulk gap at $\mu=-1$ vanishes slower with $L$ ($\sim L^{-(\alpha-1)}$) as compared to the bulk gap at $\mu=1$ ($\sim L^{-1}$). 
This implies that at $\mu=-1$, the system is farther away from criticality than that at $\mu=1$ for a finite system size $L$. Consequently, the DEE shows a much smoother transition at $\mu=-1$ than at $\mu=1$ for finite system sizes (see Fig.~\ref{fig_mu_SD_alpha_2.0}), when $1<\alpha \leq 2$.

\section{Variation of the bulk contribution to the DEE with $\alpha$}\label{sec_bulk_sd}
\begin{figure*}
	\centering
	\subfigure[]{
		\includegraphics[width=8.5cm,height=6.5cm]{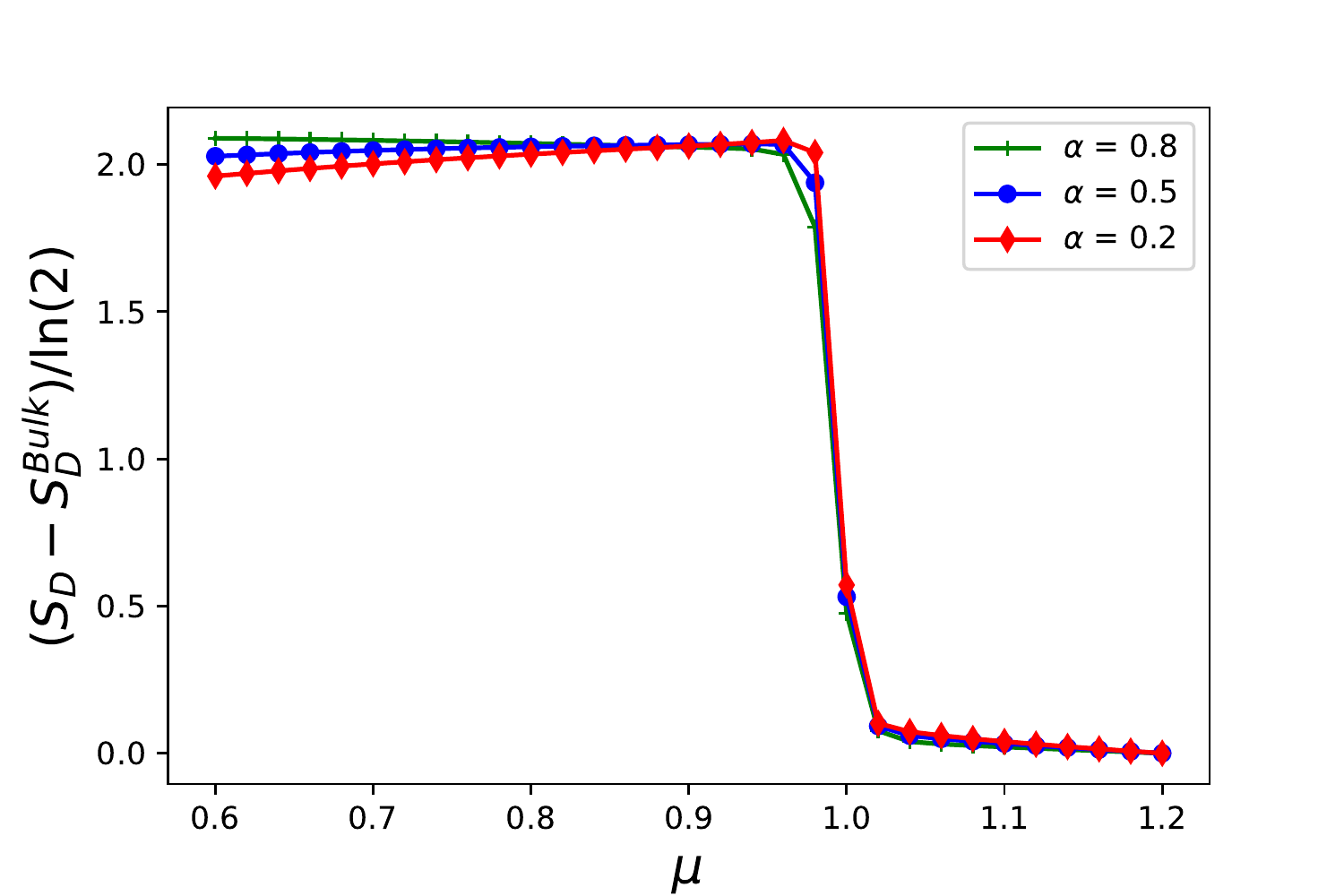}
		\label{fig_mu_sd_edge}}	
	\centering
	\subfigure[]{
		\includegraphics[width=8.5cm,height=6.5cm]{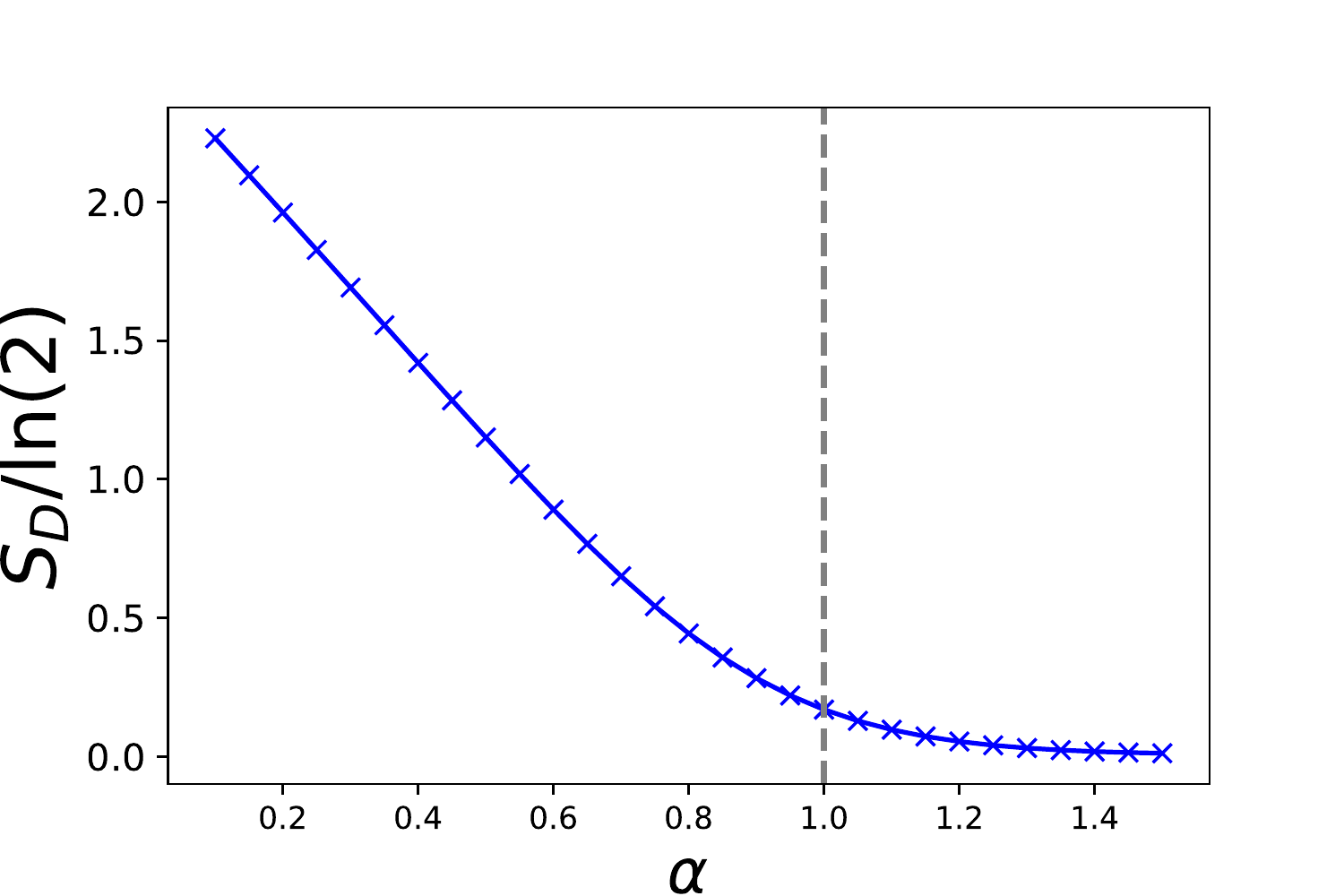}
		\label{fig_alpha_sd_fitting}}
	\caption{(a) $\left( S_{D} - S_{D}^{Bulk} \right)$ as a function of $\mu$ for different values of $\alpha <1$, close to $\mu=1$. The quantity $\left( S_{D} - S_{D}^{Bulk} \right)$ takes the value $2 \ln(2)$ for $\mu \to 1^{-}$ and $0$ for $\mu \to 1^{+}$ for all values of $\alpha<1$. (b) $S_{D}$ (i.e., bulk contribution to $S_{D}$) at $\mu=1.05$ as a function of $\alpha$. For $\alpha<1$, the bulk contribution to $S_{D}$ close to $\mu=1$ decreases with the increase of $\alpha$ and quickly approaches zero for $\alpha \gtrapprox 1$. The other relevant parameters in both the figures are $\Delta=1$, $L=100$, $L_{D}=25$.}\label{fig_sd_bulk}
\end{figure*}

In this section, we briefly discuss the bulk contribution of the DEE and its variation with $\alpha$. In the long range interacting Kitaev chain with $\alpha<1$, any finite contribution to the DEE for $\mu>1$ (trivial phase) must come from only the bulk states. Thus, in the limit $\mu \to 1^{+}$, we can write,
\begin{equation}\label{eq_bulk_sd_g}
S_{D} (\mu \to 1^{+}) = S_{D}^{Bulk} (\mu \to 1^{+}),
\end{equation}
where $S_{D}^{Bulk}$ is the contribution of the bulk states to the DEE. If we assume the continuity of the bulk contribution at $\mu=1$, then we can write,
\begin{equation}\label{eq_bulk_cont}
S_{D}^{Bulk} (\mu \to 1^{-}) = S_{D}^{Bulk} (\mu \to 1^{+}).
\end{equation}
Using Eq.~\eqref{eq_bulk_sd_g} and Eq.~\eqref{eq_bulk_cont}, we get,
\begin{multline}\label{eq_sd_less}
S_{D} (\mu \to 1^{-}) - S_{D}^{Bulk} (\mu \to 1^{-}) \\
=  S_{D} (\mu \to 1^{-}) - S_{D} (\mu \to 1^{+}).
\end{multline}
In Fig.~\ref{fig_mu_sd_edge}, the quantity $\left( S_{D} - S_{D}^{Bulk} \right)$ is plotted as a function of $\mu$ for $\alpha<1$, close to $\mu=1$. The quantity $\left( S_{D} - S_{D}^{Bulk} \right)$ takes the value $2\ln(2)$ in the limit $\mu \to 1^{-}$ for all values of $\alpha<1$. Thus, from Eq.~\eqref{eq_sd_less} and Fig.~\ref{fig_mu_sd_edge}, it is clear that 
\begin{equation}\label{eq_sd_jump}
S_{D} (\mu \to 1^{-}) - S_{D} (\mu \to 1^{+}) \approx 2 \ln(2).
\end{equation}
This implies that the jump of the DEE at the critical point $\mu=1$ is an integer multiple of $\ln(2)$ (i.e., discontinuous, quantized jump) even for the long range interacting Kitaev chain with $\alpha<1$. Also, this indicates that the contribution of the DEE coming from the bulk states is like a background value of the DEE.\\

In Fig.~\ref{fig_alpha_sd_fitting}, the DEE for $\mu \to 1^{+}$ (i.e., the bulk contribution to the DEE, see Eq.\eqref{eq_bulk_sd_g}) is plotted as function of $\alpha$. 
From this plot, it can be observed that, for $\alpha<1$, the bulk contribution of the DEE close to $\mu=1$ decreases with increasing $\alpha$ and approaches zero for $\alpha \gtrapprox 1$.\\

\section{Variation of the DEE with $L_{D}$}\label{sec_sd_ld}
\begin{figure*}
	\centering
	\subfigure[]{
		\includegraphics[width=0.45\textwidth]{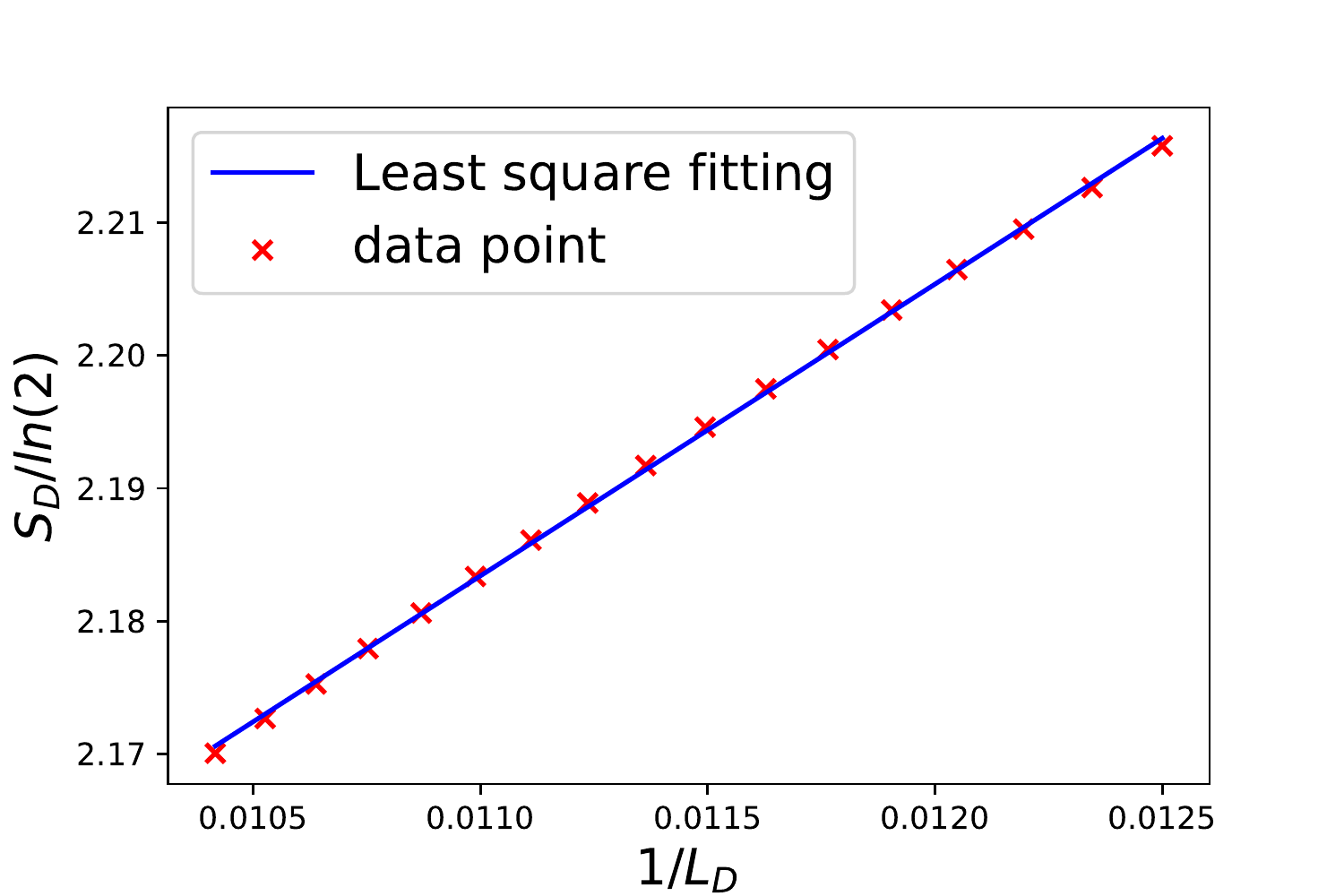}
		\label{fig_ld_sd_l}}	
	\centering
	\subfigure[]{
		\includegraphics[width=0.45\textwidth]{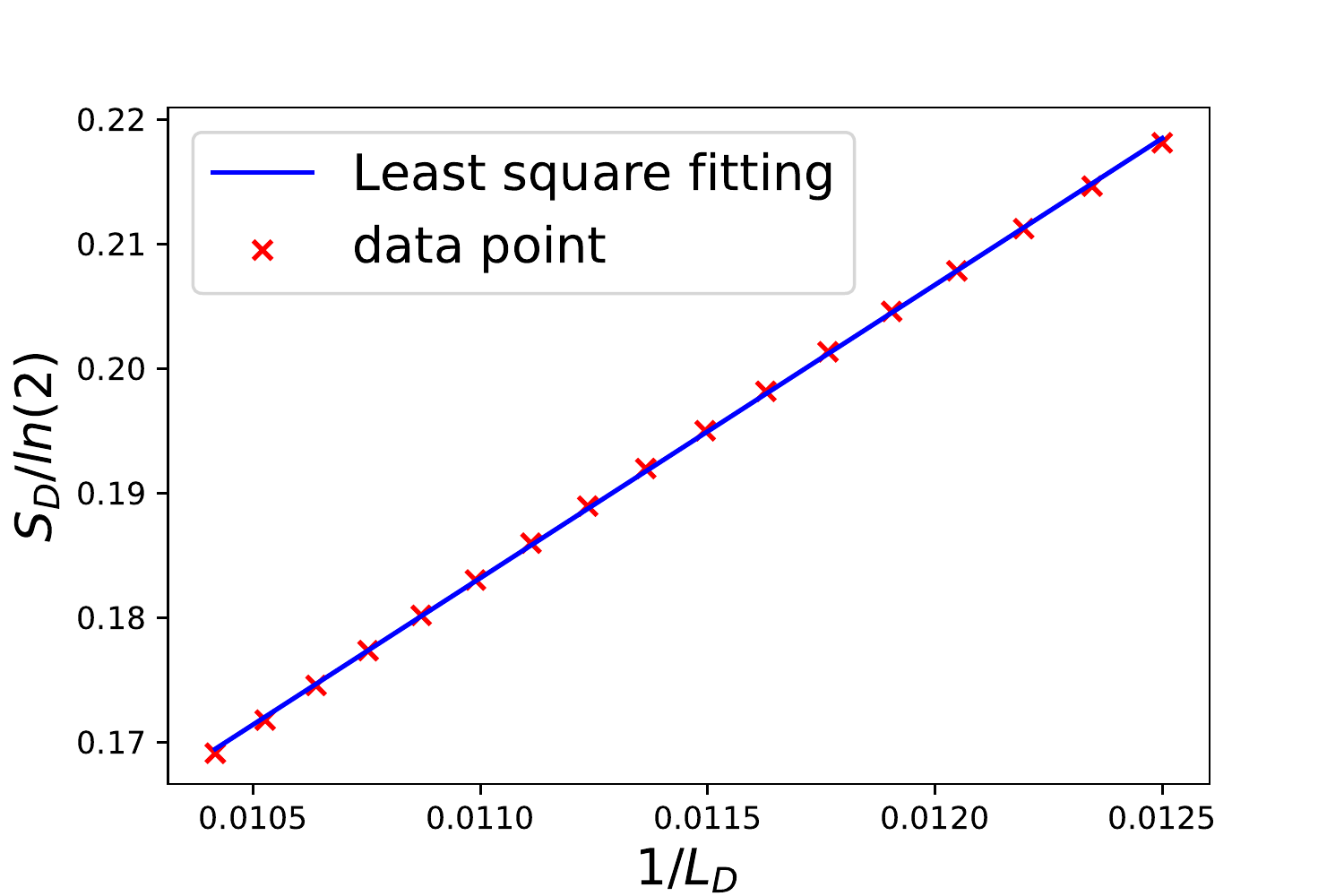}
		\label{fig_ld_sd_g}}
	\caption{$S_{D}/\ln(2)$ as a function of $1/L_{D}$ in the (a)  topologically non-trivial phase and near the critical point ($\mu=0.95$) and (b) topologically trivial phase and near the critical point ($\mu=1.01$). The other relevant parameters in both the figures are $L=200$ and $\alpha=0.9$. The DEE decreases with increasing $L_D$ for both $\mu \to 1^{-}$ and $\mu \to 1^{+}$.  From the linear fit of the plots, it can be seen that the intercepts are $b_{1}\approx 1.96$ and $b_{2}\approx 0.08$, respectively. Thus, in the limit $L_{D} \to \infty$ for thermodynamically large system, $S_{D}$ approaches  $2\ln(2)$ and $0$ for $\mu \to 1^{-}$ and $\mu \to 1^{+}$, respectively.} \label{fig_ld_SD_gl}
\end{figure*}

\begin{figure}
	\centering
		\includegraphics[width=0.45\textwidth]{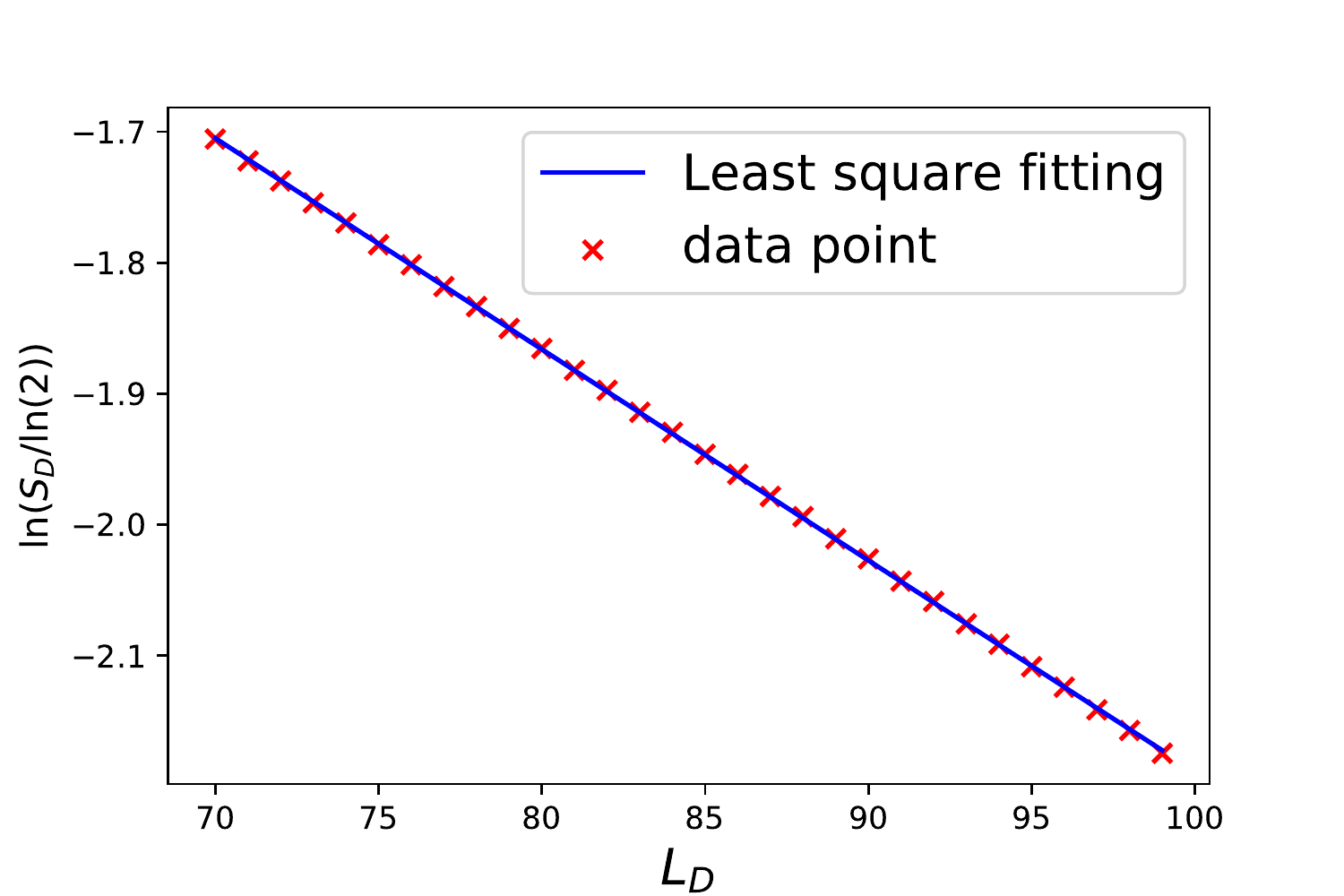}
	\caption{$\ln(S_{D}/\ln(2))$ as a function of $L_{D}$ for $\mu=1.5$ (topologically trivial phase and far away from the critical point) . The other relevant parameters in the figure are $L=200$ and $\alpha=0.9$. The DEE decreases exponentially with increasing $L_D$ in this case.} \label{fig_ld_sd_log_g}
\end{figure}

In this section, we discuss the variation of the DEE with $L_{D}$ near the critical point considering both the topologically non-trivial and trivial phase for a long range interacting Kitaev chain with $\alpha<1$. In Fig.~\ref{fig_ld_sd_l} and Fig.~\ref{fig_ld_sd_g}, we have plotted $S_{D}/\ln(2)$ for $\mu \to 1^{-}$ and $\mu \to 1^{+}$ against $1/L_{D}$. From the linear fit of the plots, it can be seen that the intercepts are $b_{1}\approx1.96$ and $b_{1}\approx 0.08$, respectively. This implies,
\begin{subequations}\label{eq_sd_fit_near}
\begin{equation}\label{eq_sd_near_l}
\frac{S_{D} (\mu \to 1^{-})}{\ln(2)} \sim \frac{a_{1}}{L_{D}} + b_{1},
\end{equation}
\begin{equation}\label{eq_sd_near_g}
\frac{S_{D} (\mu \to 1^{+})}{\ln(2)} \sim \frac{a_{2}}{L_{D}} + b_{2},
\end{equation}
\end{subequations}
where $a_1$ and $a_2$ are constants. Thus, as one approaches the critical point along the topologically non-trivial phase, i.e., $\mu\to 1^{-}$), the jump in the DEE approaches $2\ln (2)$ as $L_{D} \to \infty$, for thermodynamically large systems. It can also be seen that, by approaching the critical point along the trivial phase, i.e., $\mu\to 1^{+}$, the DEE approaches zero as $L_{D} \to \infty$. Thus, we infer that the discontinuous jumps in the DEE at critical points, indeed become quantised with increasing partition size even for long range interacting systems.\\

It can be observed from Fig.~\ref{fig_ld_sd_log_g} that the plot of $\ln (S_{D}/\ln(2))$, far away from the critical point, in the topologically trivial phase (i.e., $\mu>1$) with $L_{D}$ is linear with the slope $-c_{1}$ (where $c_{1} > 0$) with all other parameters (i.e., $\alpha, \mu$ and $L$) fixed. Thus, for the trivial phase ($\mu>1$) and away from the critical point $\mu=1$, we obtain,
\begin{align}\label{eqn_sd_logfit_l}
\ln (\frac{S_{D}}{\ln(2)}) = -c_{1} L_{D} + c_{2} , \nonumber \\
\implies \frac{S_{D}}{\ln(2)} \sim \exp \left( - \frac{L_{D}}{\lambda} \right) ,
\end{align}
for some constant $c_{2}$, where $\lambda=1/c_{1}$ is a function of $\alpha$, $\mu$ and $L$. 
Thus, we infer that the DEE diminishes exponentially with the length $L_{D}$ of the disconnected partition in the trivial phase and far away from the critical point $\mu=1$, with the length-scale $\lambda$ characteristic to the long-range interacting system.



\section{Variation of DEE with the strength of superconducting pairing ($\Delta$) in short range and long range interacting Kitaev chain}\label{delta_dee}
\begin{figure*}
	\centering
	\subfigure[]{
		\includegraphics[width=0.45\textwidth]{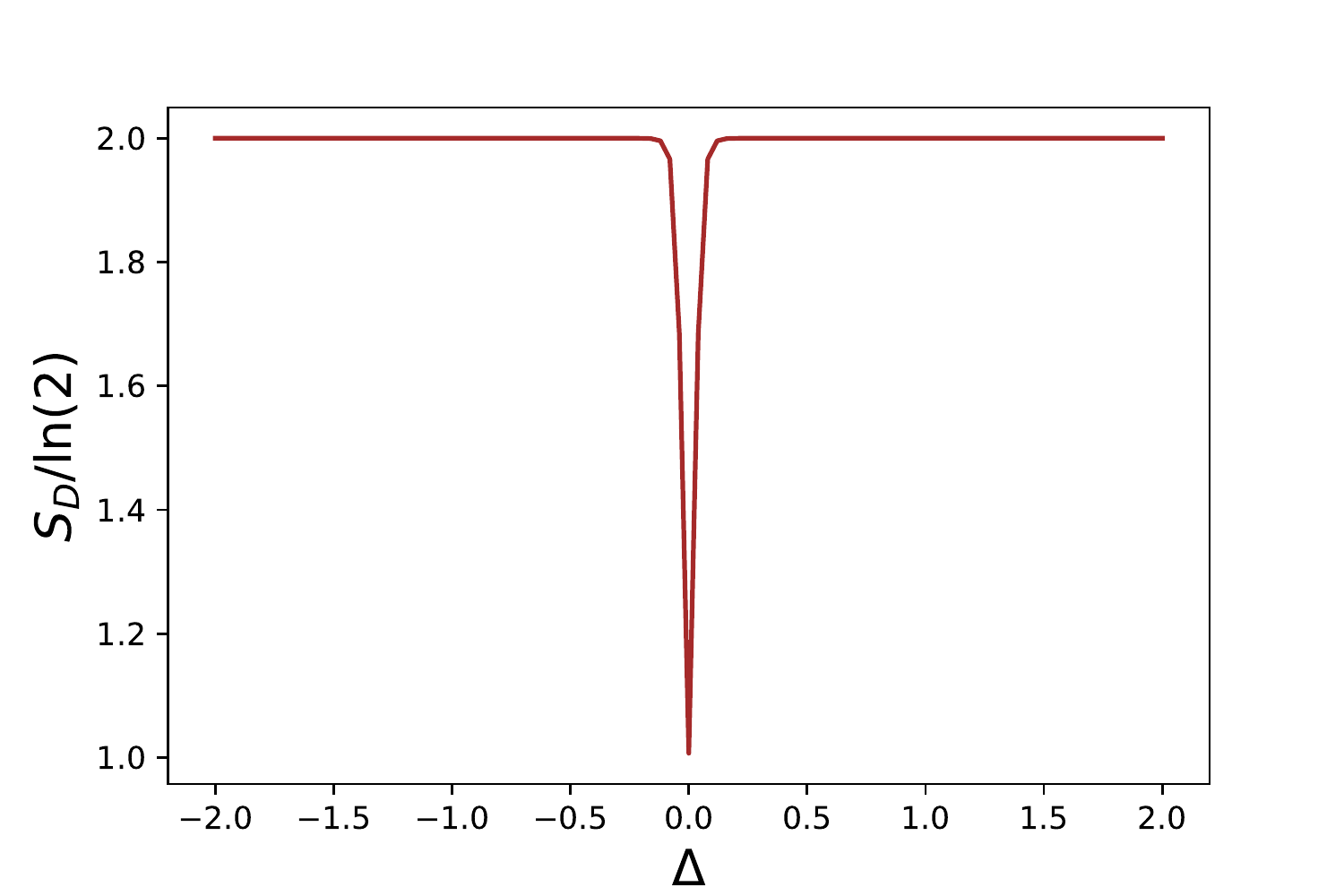}
		\label{fig_delta_sd_srk}}	
	\centering
	\subfigure[]{
		\includegraphics[width=0.45\textwidth]{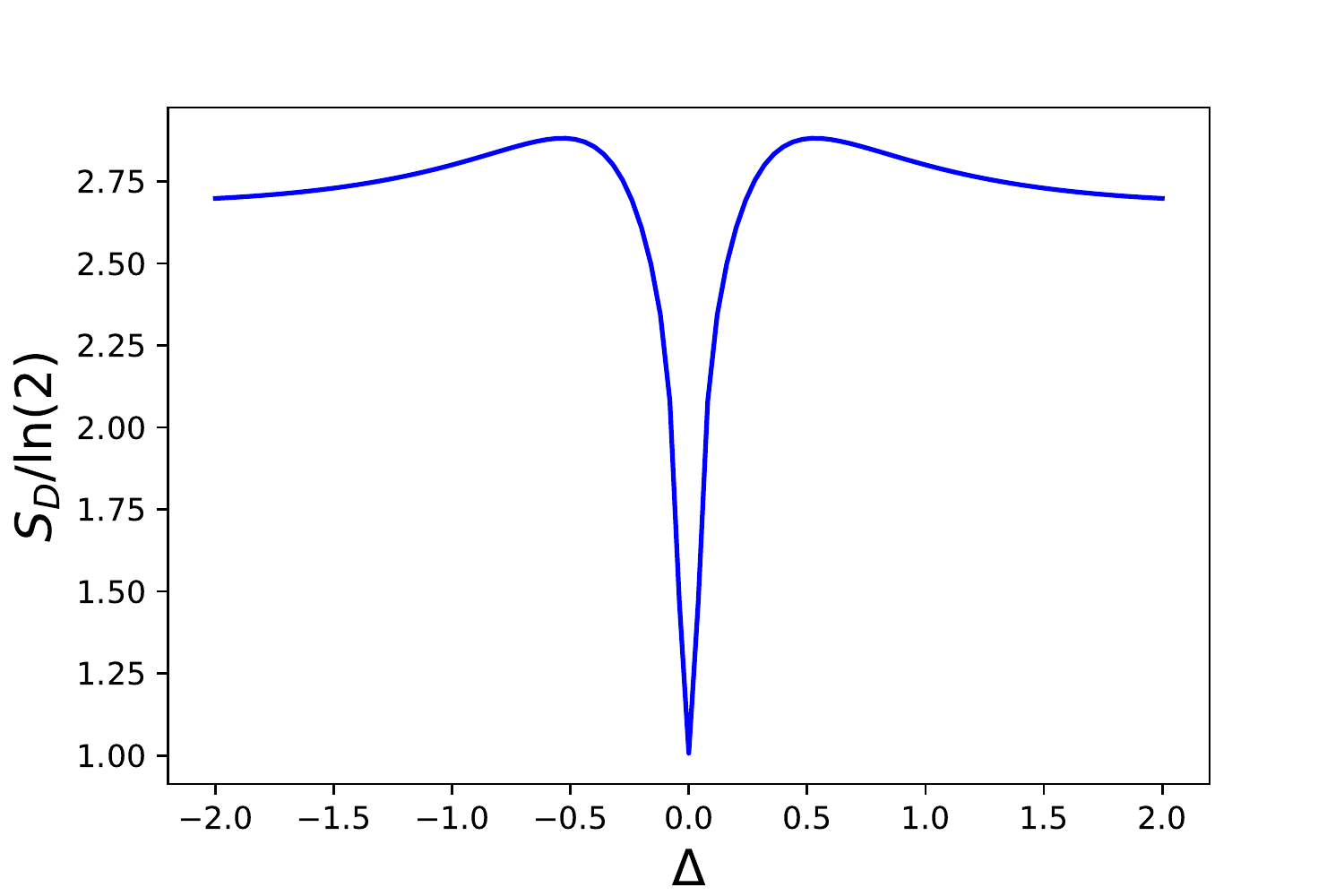}
		\label{fig_delta_sd_lrk}}
	\caption{The DEE as a function of $\Delta$ in the (a) short range limit and (b) long range scenario with $\alpha=0.5$. The other relevant parameters chosen are $\mu=0$, $L=100$ and  $L_{D} =25$. In both the cases, the DEE exhibits a jump at $\Delta=0$.}\label{fig_sd_delta}
\end{figure*}
Variations of the DEE as a function of $\Delta$ for both the short range and long range interacting Kitaev chain are shown in Fig.~\ref{fig_delta_sd_srk} and Fig.~\ref{fig_delta_sd_lrk}. From both the figures, it is clear that the DEE exhibits a jump at $\Delta=0$ for the short range as well as the long range interacting Kitaev chain. As the number of Majorana modes does not change across the critical point $\Delta=0$ for the short range interacting Kitaev chain, the DEE saturates to the same value $2\ln(2)$ on both sides of the critical point $\Delta=0$ in the topologically non-trivial phase (i.e., $-1<\mu<1$).

\section{Calculation of effective range of interactions appearing in $H_{\rm eff}$ for short range Kitaev chain}\label{app_H_eff}
Let us consider the following forms of the Hamiltonians $H_{i}$ and $H_{f}$ :\\
\begin{multline}\label{eqn_SRK_H_i}
H_{i} = \sum_{n=1}^{L-1} -( c_{n}^{\dagger} c_{n+1} + c_{n+1}^{\dagger} c_{n} ) - \mu_{i} \sum_{n=1}^{L} ( 2 c_{n}^{\dagger} c_{n} -1 ) \\ + \sum_{n=1}^{L-1} \Delta ( c_{n} c_{n+1} + c_{n+1}^{\dagger} c_{n}^{\dagger}),
\end{multline}
\begin{multline}\label{eqn_SRK_H_f}
H_{f} = \sum_{n=1}^{L-1} -( c_{n}^{\dagger} c_{n+1} + c_{n+1}^{\dagger} c_{n} ) - \mu_{f} \sum_{n=1}^{L} ( 2 c_{n}^{\dagger} c_{n} -1 ) \\ + \sum_{n=1}^{L-1} \Delta ( c_{n} c_{n+1} + c_{n+1}^{\dagger} c_{n}^{\dagger}).
\end{multline}
Now, using Eq.~\eqref{eqn_SRK_H_i}, Eq.~\eqref{eqn_SRK_H_f} and the usual anti-commutation relations of Fermionic annihilation and creation operators, we explicitly calculate $K_{r} (H_{f},H_{i})=[H_{f},[H_{f},[....,[H_{f},H_{i}]....]]]$ (see Sec.~\ref{sec_time_evolution_SD}), for $r =$ integer. Thus, we obtain,
\begin{equation}\label{eqn_K1_Hf_Hi}
K_{1} (H_{f}, H_{i}) = 4 \Delta (\mu_{f}-\mu_{i}) \sum_{n=1}^{L-1} (c_{n} c_{n+1} - c_{n+1}^{\dagger} c_{n}^{\dagger}) ,
\end{equation}
\begin{multline}\label{eqn_K2_Hf_Hi}
K_{2} (H_{f}, H_{i}) = 8 \Delta (\mu_{f} - \mu_{i}) \sum_{n=1}^{L-2} ( c_{n} c_{n+2} + c_{n+2}^{\dagger} c_{n}^{\dagger} - \Delta c_{n}^{\dagger} c_{n+2}\\ -  \Delta c_{n+2}^{\dagger} c_{n}) + 16 \mu_{f} \Delta (\mu_{f} - \mu_{i}) \sum_{n=1}^{L-1} (  c_{n} c_{n+1} +  c_{n+1}^{\dagger} c_{n}^{\dagger}) \\ +  8 \Delta ^{2} (\mu_{f} - \mu_{i}) \sum_{n=1}^{L} (2 c_{n}^{\dagger} c_{n} - 1) .
\end{multline}

Thus, it is easy to observe that the maximum range of interactions appearing in $K_{r}(H_{f}, H_{i})$  is $r$, for all $r =$ integer. In general, for any $r =$ integer, it can be written that,
\begin{multline}\label{eqn_K_2r_Hf_Hi}
K_{2 r} (H_{f}, H_{i}) = \sum_{n=1}^{L-1} \sum_{l=1}^{2 r} (\beta_{2 r,l}c_{n}^{\dagger} c_{n+l} + \delta_{2 r,l} c_{n} c_{n+l} + \rm {H.c.}) \\ + \lambda_{2 r} \sum_{n=1}^{L} (2 c_{n}^{\dagger} c_{n} -1) ,
\end{multline}
\begin{equation}\label{eqn_K_2r+1_Hf_Hi}
K_{2 r + 1} (H_{f}, H_{i}) = \sum_{n=1}^{L-1} \sum_{l=1}^{2 r +1} (\delta_{2 r +1, l} c_{n} c_{n+l} - \rm {H.c.}) ,
\end{equation}
where $\beta_{2 r,l}$, $\delta_{2 r,l}$, $\lambda_{2 r}$ and $\delta_{2 r + 1 ,l}$  are the functions of $\Delta$, $\mu_{i}$ and $\mu_{f}$. It can also be seen that 
\begin{equation}\label{eqn_K_r_dagger}
K^{\dagger}_{r}(H_{f},H_{i}) = (-1)^{r} K_{r}(H_{f},H_{i}) ,
\end{equation}
for all $r =$ integer, which follows from the fact that the effective Hamiltonian $H_{\rm eff}$ is an Hermitian operator.\\

Putting $t=0$ in Eq.~\eqref{eqn_H_eff_series} (see Sec.~\ref{sec_time_evolution_SD}), we get, $H_{\rm eff} (t = 0) = H_{i}$. As we have assumed that $H_{i}$ consists of only the short ranged (only the nearest neighbours are interacting) interactions, the maximum range of interactions appearing in $H_{\rm eff} (t=0)$, $r_{\rm max}(t=0) =1$.
At very small $t$ (i.e., $t << 1$), the terms of the order $t^2$ and higher can be neglected and we have,
\begin{equation}\label{eqn_Heff_t1}
H_{\rm eff}(t)= H_{i} -i t K_{1} (H_{f},H_{i}) + \mathcal{O}(t^2),
\end{equation}
which implies that the maximum effective range of interaction, $r_{\rm max}(t)=1$, as maximum range of interaction appearing in $K_{1} (H_{f},H_{i})$ is $1$.
As the time ($t$) increases, we need to consider the terms with higher orders in $t$ of Eq.~\eqref{eqn_H_eff_series}. As the maximum range of interactions appearing in $K_{r} (H_{f},H_{i})$ is $r$, the effective range ($ r_{\rm max} (t)$) of interactions appearing in $H_{\rm eff} (t)$ is much larger (i.e., $r_{\rm max} (t) >> 1$) after sufficiently large time (i.e., $t >> 1$). Thus, we can say that the effective range of interaction $r_{\rm max}(t)$ increases with time.

\section{Time evolution of DEE following sudden quenches of the strength of superconducting pairing ($\Delta$) and the exponent $\alpha$ of the power law interaction}\label{dee_time_delta}
\begin{figure*}
	\centering
	\subfigure[]{
		\includegraphics[width=0.45\textwidth]{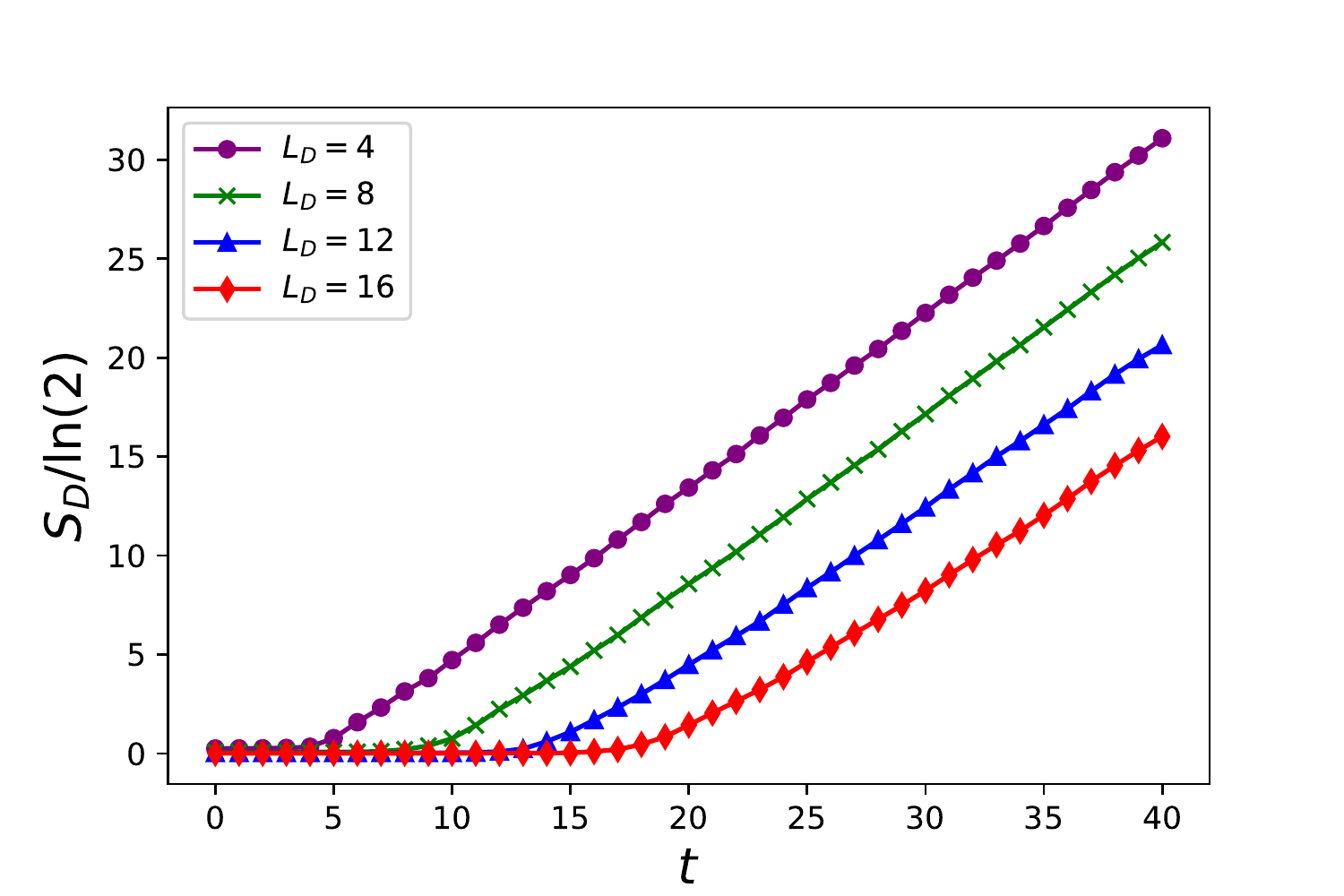}
		\label{fig_t_SD_delta_ld}}	
	\centering
	\subfigure[]{
		\includegraphics[width=0.45\textwidth]{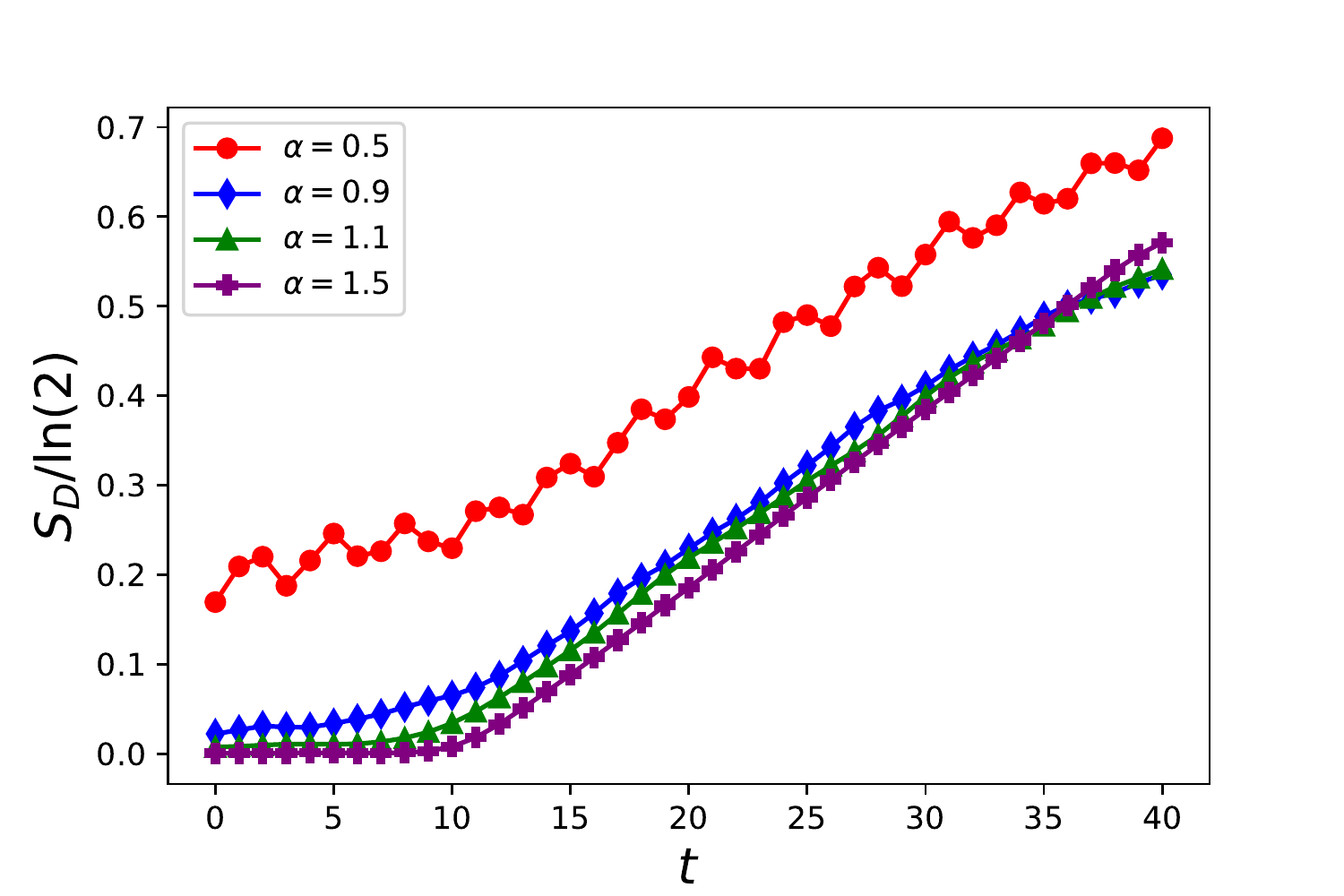}
		\label{fig_t_SD_delta_alpha}}
	\caption{(a) $S_{D}$ as a function of time ($t$) for different values of $L_{D}$ for short range interacting Kitaev chain with $L=40$, $\Delta_{i}=10$ and $\Delta_{f} = 1$, $\mu = 2$, (b) $S_{D}$ as a function of time for different values of $\alpha$ for long range interacting Kitaev chain with $L=40$, $L_{D}=10$, $\Delta_{i}=10$ and $\Delta_{f}=1$, $\mu = 100$. Here, $\Delta_{i}$ and $\Delta_{f}$ denote the strengths of superconducting pairing before and after the quench, respectively.} \label{fig_t_SD_delta}
\end{figure*}

\begin{figure}
	\centering
		\includegraphics[width=0.45\textwidth]{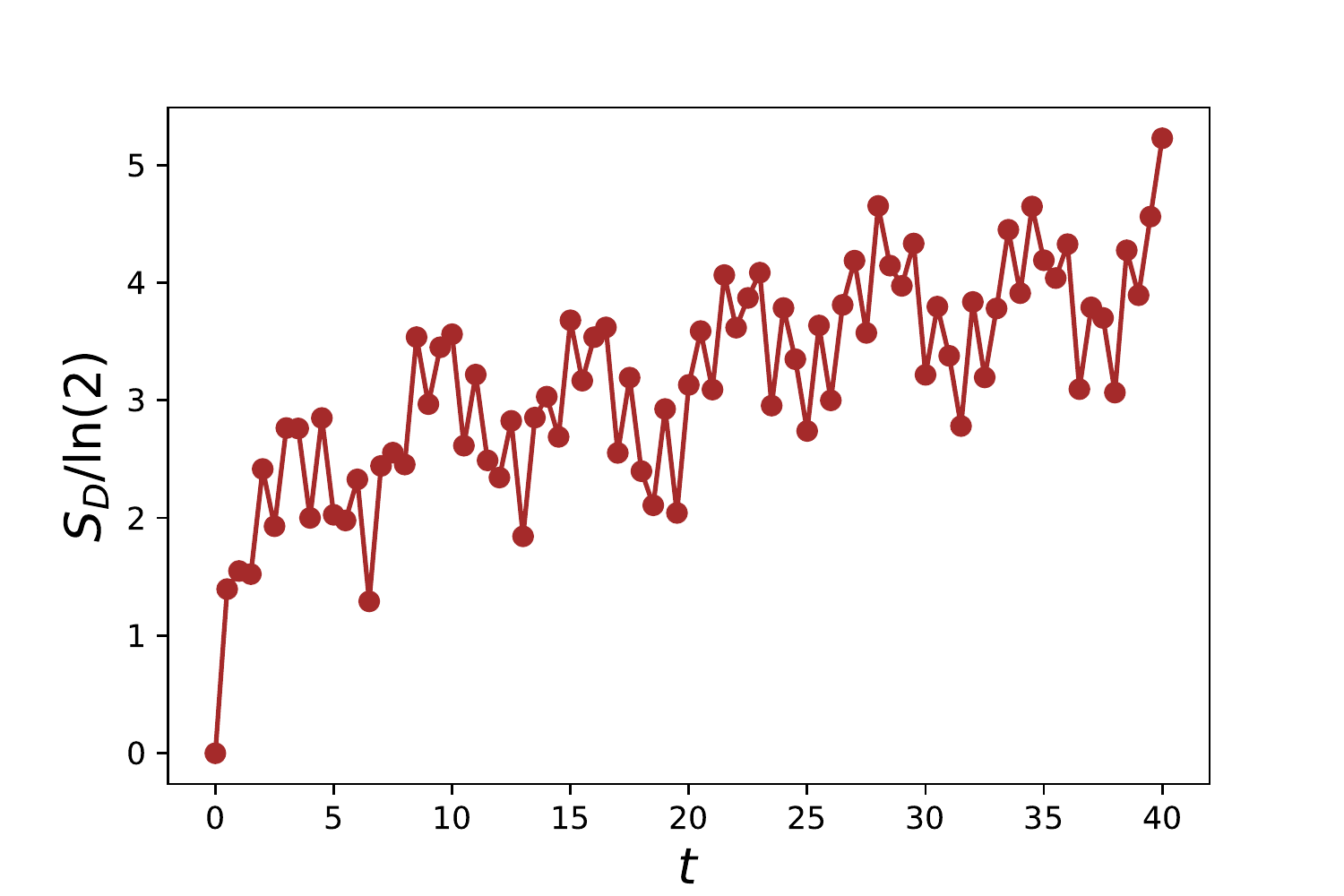}
	\centering
	\caption{$S_{D}$ as a function of time ($t$) for Kitaev chain with $L=40$, $\alpha_{i}=5$ and $\alpha_{f} = 0.5$, $\mu = 2$, $\Delta=1$. Here, $\alpha_{i}$ and $\alpha_{f}$ denote the the exponent determining the range of power-law decaying interaction before and after the quench, respectively.} \label{fig_t_SD_alpha_vary}
\end{figure}

In this section, we discuss the time evolution of the DEE following sudden quenches of the strength of the superconducting pairing ($\Delta$) and the exponent $\alpha$ of the power law interaction. From Fig.~\ref{fig_t_SD_delta_ld}, it is clear that  $S_{D}$ remains invariant upto a non-zero critical time $t_{c}$, proportional to the length $L_{D}$ of the disconnected partition, following a sudden quench of the parameter $\Delta$ within the same topological phase (keeping initial and final Hamiltonians $H_{i}$ and $H_{f}$ topologically equivalent) in a short range interacting Kitaev chain with fixed finite length $L$. Also, from Fig.\ref{fig_t_SD_delta_alpha}, we observe that the critical time $t_{c}$ turns out to be zero for long range interacting Kitaev chain. Thus, we conclude that  in the long range interacting systems, the DEE following a sudden quench in either of the parameters $\mu$ or $\Delta$, does not remain invariant with time even when $H_{i}$ and $H_{f}$ are topologically equivalent.

We also observe the behaviour of the DEE following a sudden quench of the parameter $\alpha$. If the system is suddenly quenched from $\alpha>1$ to $\alpha<1$ (i.e., a short range interacting system is suddenly converted to a long range interacting system), then $t_{c}$ is found to be zero, as can be seen from Fig.~\ref{fig_t_SD_alpha_vary}. Therefore, we conclude that if any one of $H_{i}$ and $H_{f}$ is long-ranged, the critical time $t_{c}$ is zero. This indicates a breakdown of the topological classification through the DEE, in out of equilibrium long range interacting systems.

\end{document}